\documentclass[12pt]{article}

\usepackage{amsmath}
\usepackage[dvipdfmx]{graphicx}
\usepackage{natbib}

\usepackage{bm}

\usepackage{amsthm}
\theoremstyle{plain}
\newtheorem{prop}{Proposition}
\theoremstyle{remark}

\usepackage[hang,small,bf]{caption}
\usepackage[subrefformat=parens]{subcaption}
\usepackage[dvipsnames]{xcolor}

\begin{document}

\def\spacingset#1{\renewcommand{\baselinestretch}%
{#1}\small\normalsize} \spacingset{1}


\title{Economic variable selection}
\author{
Koji Miyawaki \\ School of Economics, Kwansei Gakuin University
\and
Steven N. MacEachern \\ Department of Statistics, The Ohio State University
}
\date{\today}
\maketitle

\begin{abstract}
Regression plays a central role in the discipline of Statistics and is the primary analytic technique in many research areas.
Variable selection is a classic and major problem for regression.
This study emphasizes the economic aspect of variable selection.
The problem is formulated in terms of the cost of predictors to be purchased for future use: only the subset of covariates used in the model will need to be purchased.
This leads to a decision-theoretic formulation of the variable selection problems that includes the cost of predictors as well as their effect.
We adopt a Bayesian perspective and propose two approaches to address uncertainty about model and model parameters.
These approaches, termed the restricted and extended approaches, lead us to rethink model averaging.
From objective or robust Bayes point of view, the former is preferred.
The proposed method is applied to three popular datasets for illustration.

\end{abstract}

\def\spacingset#1{\renewcommand{\baselinestretch}{#1}\small\normalsize} \spacingset{1}
\spacingset{1.45}


\section{Introduction}


Model selection with subsequent prediction is a classic and major problem in statistics.
In the context of regression analysis, model selection is often equated with variable selection, to be accomplished in one of many ways, including classical hypothesis test of full and reduced models (e.g., \citet{voung-89}), use of an information criterion such as AIC, BIC, or DIC (\citet{akaike-98} for a reprint of the original paper published in 1973 for AIC, \citet{schwarz-78} for BIC, and \citet{spiegelhalter-etal-02} for DIC), evaluation through some form of cross-validation (e.g., \citet{gelfand-eta-92} and \citet{gelfand-dey-94}), or Bayesian versions of tests based on the Bayes factor (\citet{kass-raftery-95}).
Subsequent prediction follows from either a separate re-fit of the data as with model selection followed by a least squares fit, or is integrated into a cohesive framework involving selection and prediction as with many of the recently-developed penalized likelihood methods.

Bayesian methods provide a distinct approach to model selection and prediction, as they are based on a cohesive modelling framework that allows one to simultaneously describe and work with their uncertainty across models and over parameters within a model.
Its main applications in model selection are the hierarchical approach (\citet{mitchell-beauchamp-88} and \citet{george-mcculloch-93}) or the stochastic search approach (\citet{hans-etal-07} and \citet{fouskakis-draper-08}).
These methods follow the usual route from prior distribution through data to posterior distribution, with inference to follow.
Model selection follows from inference designed to minimize incorrect model selection while prediction follows inference to minimize forecasting loss.
This approach separates modelling from inference, facilitating for example, \citet{barbieri-berger-04} to distinguish model selection from variable selection.
It has also led to an explosion of literature on \textit{model averaging}, such as \citet{min-zellner-93}, \citet{madigan-etal-95}, \citet{raftery-etal-97}, \citet{draper-95}, \citet{brown-etal-02}, and \citet{yu-etal-11}, whose benefits are now well-documented.

The split between modelling and inference has generated a novel approach to parsimony within Bayesian circles which may be characterized as ``fit in a large model space, make inference in a small model space (\citet{walker-gutierrez-pena-99}, \citet{maceachern-01}, and \citet{hahn-carvalho-15}).
This approach moves parsimony from modelling to inference.
It seeks to construct a model that reflects the full complexity of the problem and, if little benefit is shown for some variables (aspects of the model), to move to a simpler form as part of inference.
In this work, we explicitly bring an economic question into the mix---namely the cost of predictors---and pursue a path suggested by the decision-theoretic formulation of the model/variable selection and prediction problem in regression.
This version places our focus on two main questions:
\begin{enumerate}
\item Prediction, accounting for the cost of predictors.
      In a typical setting, predictors have costs associated with them.
      They cost money, take time to collect, take effort to model, or consume computational effort.
      These costs are real, and obtaining a slightly better prediction rule at a much higher cost or much more slowly may not be worthwhile.
\item Model uncertainty. The goal of model selection is often taken to be consistent model selection, or identification of the set of predictors with nonzero coefficients in the regression model.
      The economic formulation of the prediction problem suggests that a slightly inferior (in the traditional sense) model may provide a better model for practical use.
      This suggests a re-examination of the role of consistency in model selection.
      See also \citet{clyde-george-04} for recent approaches to model uncertainty.
\end{enumerate}

Consideration of the cost of predictors and formulation of model selection as a decision problem has appeared in the literature.
\citet{lindley-68} argues forcefully for Bayesian methods and for a full decision-theoretic formulation of the problem.
Further authors have mentioned this issue (see, for example, \citet{brown-etal-99} and \citet{hahn-carvalho-15}).
\citet{fouskakis-etal-09a} and \citet{fouskakis-etal-09b} also take the cost into account in different directions.

In this work, we seek to reconcile economic considerations with current practice in Bayesian model selection and model averaging.
We find that this perspective provides strong commentary on current practice, we present several reasons to believe that current practice is generally reasonable, and we identify settings where improvements can be made.
In all, we find that Bayesian model averaging (BMA) is a valuable technique but that care should be taken to its implementation.

This paper is organized as follows.
The next section laid out our methodology, including choice of two approaches (Subsections \ref{subsec:Two approaches} and \ref{subsec:Choice of approach}).
Three data sets are used to illustrate our method in Section \ref{sec:Illustrative examples}.
Section \ref{sec:Discussion} points to future directions and concludes the paper.


\section{Economic variable selection}
\label{sec:Economic variable selection}


\subsection{Normal linear model with $g$ prior}
\label{subsec:Normal linear model with g prior}


Suppose we have the response $Y_{i}$ and $p$ potential predictors $\bm{x}_{i}$ for each $i = 1, \dots, n$ observation ($p < n-1$).
All predictors are standardized.
A subset of $p$ predictors is indexed by $\gamma$ and is denoted by $\bm{x}_{i, \gamma}$, where $\gamma = (\gamma_{1}, \dots, \gamma_{p})^{\prime}$ is a vector of ones and zeros to indicate which predictors are in the model.
When the $j$-th predictor is in the model $\gamma$, the $j$-th element $\gamma_{j}$ is set to one, and it is zero otherwise ($j = 1, \dots, p$).
Let $p_{\gamma}$ be the dimension of $\bm{x}_{i, \gamma}$ and $\Gamma$ be the set of all possible $\gamma$s.

This paper focuses on the model, details of which are given below.
For $\gamma \in \Gamma$ and $i = 1, \dots, n$,
\begin{align}
Y_{i} = \beta_{0} + \bm{x}_{i, \gamma}^{\prime} \bm{\beta}_{1, \gamma} + \epsilon_{i}, \notag
\end{align}
where the error term $\epsilon_{i}$ independently and identically follows the normal distribution with mean $0$ and variance $\sigma^{2}$, i.e., $\epsilon_{i} \sim N \left( 0, \sigma^{2} \right)$.
Because the model is indexed by $\gamma$ and $\gamma$ is associated with $\bm{x}_{i}$, every model includes the intercept.
The design matrix $\bm{X}_{\gamma} = (\bm{x}_{1, \gamma}, \dots, \bm{x}_{n, \gamma})^{\prime}$ is assumed to be of full column rank, which is satisfied in all examples in Section \ref{sec:Illustrative examples}.

Model parameters are $\bm{\phi}_{\gamma} = ( \beta_{0}, \bm{\beta}_{1, \gamma}, \sigma^{2} )$.
The subscript for $\beta_{0}$ and $\sigma^{2}$ is suppressed because they are commonly used in all models.
In particular, because predictors are standardized, $\beta_{0}$ is interpreted as mean of the response variable.
Thus, it is reasonable to assume its prior knowledge to be same and noninformative across all models.
While the error variance does not have such interpretation, we assume in a similar manner because it is a nuisance parameter.
For other parameters, proper prior distributions are assumed.
There are $2^{p}$ possible models ($| \Gamma | = 2^{p}$).
Prior distribution on this model space is assumed as noninformative because we have little information on which models are better in general.
In summary, the following prior distributions are assumed:
\begin{align*}
\pi \left( \beta_{0} \right) &\propto 1,
\quad
\bm{\beta}_{1, \gamma} \sim N_{p_{\gamma}} \left\{ \bm{0}, g \sigma^{2} \left( \sum_{i = 1}^{n} \bm{x}_{i, \gamma} \bm{x}_{i, \gamma}^{\prime} \right)^{-1} \right\}, \notag \\
\pi \left( \sigma^{2} \right) &\propto \sigma^{-2},
\quad
\pi \left( \gamma \right) = \prod_{j = 1}^{k} s^{\gamma_{j}} \left( 1 - s \right)^{1 - \gamma_{j}},
\end{align*}
where $g$ and $s$ are known constants, $N_{k} (\bm{\mu}, \bm{\Sigma})$ is the $k$-dimensional multivariate normal distribution with mean vector $\bm{\mu}$ and variance covariance matrix $\bm{\Sigma}$.

The prior for slope coefficients ($\bm{\beta}_{1, \gamma}$) is called the $g$ prior (see \citet{zellner-86} and \citet{zellner-siow-80}).
The constant $g$ is set equal to the number of observations $n$, which is recommended by \citet{fernandez-etal-01} when the number of observations is greater than the squared number of predictors.
This prior specification is often called a class of benchmark priors.
We choose it for operational simplicity, but other choices are applicable.

When $g = n$, the resulting $g$-prior has the unit information, where information in the prior is equal to information from one observation with respect to the Fisher information matrix.
Such a prior is proposed by \citet{kass-wasserman-95} to show the relationship between the Bayes factor and BIC.
When $g$ is equal to the squared number of predictors, the resulting $g$-prior satisfies the risk inflated criterion (RIC), proposed by \citet{foster-george-94} in relation to the minimaxity.

It is possible to assume a hyperprior on $g$.
Popular choices are the hyper-$g$ and hyper-$g/n$ priors proposed by \citet{liang-etal-08} and the robust prior proposed by \citet{bayarri-etal-12}.
In Subsection \ref{subsec:Prior sensitivity}, we examine the prior sensitivity under the restricted approach (see Subsection \ref{subsec:Two approaches}) and empirically show it is robust to the prior choice between the above specification and other choices mentioned above.

The benchmark Beta prior by \citet{ley-steel-12} and the block hyper-$g$ prior by \citet{som-etal-15} would be other choices.
See also \citet{ley-steel-12} for other hyperpriors including \citet{maruyama-george-11} as well as their performances on the numerical and empirical dataset.
Priors from the objective perspective are extensively reviewed in \citet{consonni-etal-18}.


Due to the lack of knowledge about models, we set $s = 0.5$, leading to the uniform prior over models.
Other specifications of the prior model probability are found in, e.g., \citet{steel-19}.

Under above specifications, the (marginal) posterior of $(\beta_{0}, \bm{\beta}_{1, \gamma})$ is a generalized multivariate $t$ distribution and the posterior model probability is proportional to the marginal likelihood, $m (\bm{y} \mid \bm{X}_{\gamma})$, details of which are given in Appendix \ref{sec:Posterior, marginal likelihood, and loss}.
%




\subsection{Predictive loss}


Predictive regression modelling is often formulated as a decision problem, and it can be argued that this formulation underlies BMA.
The traditional formulation of the problem is driven by a predictive loss of the form $L(y, \widehat{y ( \bm{x} )}) = (y - \widehat{y ( \bm{x} )})^{2}$, where $y$ is a response to be predicted and $\widehat{y ( \bm{x} )}$ is the predicted value associated with predictors $\bm{x}$.
Using standard models and integrating over the conditional distribution of the future $y$, this loss becomes a loss taking parameter and action as arguments, namely
\begin{align}
L \left( E[Y \mid \bm{x}], \widehat{y(\bm{x})} \right) = \left( \widehat{y(\bm{x})} - E[Y \mid \bm{x}] \right)^{2} = E \left[ L \left( Y, \widehat{y(\bm{x})} \right) \mid \bm{x} \right] - V[Y \mid \bm{x}].
\label{eqn:loss_one}
\end{align}
The variance term in equation \eqref{eqn:loss_one} does not depend on the decision rule and hence can be ignored in determination of the optimal rule.
Under the normal linear regression model described in the previous subsection, this predictive loss is estimated by equation $\eqref{eq:spe given model}$ in Appendix \ref{sec:Posterior, marginal likelihood, and loss}.

Our focus is on Bayesian procedures, and the Bayes rule is the Bayesian's optimal decision rule.
It is typically constructed from the Bayesian posterior conditional viewpoint (\citet{berger-85}), by moving from prior distribution to posterior distribution and then choosing the action to minimize posterior expected (against the posterior distribution) loss.

BMA focuses on the setting where the prior distribution cuts across slices of the parameter space that are naturally described as models.
In the case of linear regression with a set of $p$ potential predictors, across the entirety of ${\cal R}^p$ for the predictors' regression coefficients.
A model is defined by the set of non-zero regression coefficients, and the set of $2^p$ potential models partition ${\cal R}^p$.
The prior distribution on these coefficients is of mixed form.
It typically assigns positive probability to each element of the partition---that is, to each subset of ${\cal R}^p$ that corresponds to a model.
The support of the prior is the entirety of each element, leading to an overall support of all of ${\cal R}^p$.
The prior distributions that underlie BMA are thus seen to be of slightly non-standard form, but they are prior distributions.

From this perspective, BMA follows directly as a standard Bayesian procedure.
Pass from prior distribution to posterior distribution via Bayes Theorem.
Once arriving at the posterior distribution, find the optimal (posterior) action.
In this case, the action happens to be expressed as a summary of the model-averaged posterior distribution, or, for squared-error loss, as model-averaged posterior predictive means.
BMA is nothing more (nor less) than sound application of Bayes Theorem and choice of an appropriate action.
As such, it inherits all of the optimality properties of Bayesian inference.




\subsection{Decision with cost}


In the variable selection problem, the basic decision involves two sets of possibly overlapping and possibly null sets of predictors.
The analyst must decide which to purchase, knowing the state of nature.
For this decision, it is important to consider their costs as well as their predictive adequacies.

To this end, following \citet{lindley-68}, we modify the original predictive loss to include the cost of data acquisition, modelling and processing, including the cost to purchase information (as in credit history for a customer), a cost of time (as in the delay in obtaining results from a medical lab test), cost in processing time (as in variables that are computationally expensive in conjunction with their use in a model), or other.

More specifically, suppose we have $(Y, \bm{x})$, a single future case.
Let $c (\gamma)$ be the nonnegative cost function for the model $\gamma$ that uses the single future case.
The cost depends on the set of predictors, but it does not depend on the values of those predictors.
Without knowledge of predictors, a typical choice is a function of the number of predictors.
Section \ref{sec:Illustrative examples} provides specific forms of the cost function.

The total cost, or the negative utility, from purchasing predictors $\bm{x}$ is expressed as the sum of $E [ L ( Y, \widehat{y(\bm{x})} ) \mid \bm{x} ]$ and $c (\gamma)$.
It is better to purchase $\bm{x}_{1}$ than $\bm{x}_{2}$ if
\begin{align*}
E \left[ L \left( Y, \widehat{y(\bm{x}_{1})} \right) \mid \bm{x}_{1} \right] + c \left( \gamma_{1} \right)
<
E \left[ L \left( Y, \widehat{y(\bm{x}_{2})} \right) \mid \bm{x}_{2} \right] + c \left( \gamma_{2} \right),
\end{align*}
where $\gamma_{1}$ and $\gamma_{2}$ are models associated with $\bm{x}_{1}$ and $\bm{x}_{2}$, respectively.
In general, the best predictor to purchase is chosen by solving the minimization problem:
\begin{align*}
\min_{\gamma \in \Gamma} E \left[ L \left( Y, \widehat{y (\bm{x}_{\gamma})} \right) \mid \bm{x}_{\gamma} \right] + c \left( \gamma \right).
\end{align*}
%
If all predictors are free of charge, it is clear that the best combination of predictors is the one that minimizes the loss.

The idea of model selection (or variable selection in the normal linear regression model) as above goes back to \citet{lindley-68}.
A recent study is \citet{gelfand-ghosh-98}.
They propose the model selection criterion by using the weighted sum of losses based on the future and current data, and discuss its properties and generalizations.
The cost function in our case can be interpreted as a specific form of the loss based on the current data.
More general discussion on this utility-based approach is found in \citet{bernardo-smith-85} for example.


\subsection{Two approaches}
\label{subsec:Two approaches}


In reality, the state of nature is unknown.
Its uncertainty is specified as the form of distribution about model parameters and about models.
Let $\bm{x}$ be the $k$ purchased predictors, and let $\bm{w}$ denote the $p-k$ unpurchased predictors.
The predictors may or may not be relevant to predict the response, and we expect future data of the form $(Y, \bm{x})$ to reveal the relationship between the response and the purchased predictors.
There are two main approaches to provide forecasts for future $Y$ as a function of the future covariate ${\bm x}$.

\noindent
{\bf The restricted approach.}
The restricted approach confines us to the small world of predictors $\bm{x}$ and response $Y$.
BMA applied to this world results in model averaging across $2^{k}$ potential models, with individual predictors in $\bm{x}$ either active or not.

\noindent
{\bf The extended approach.}
The extended approach considers the large world of models determined by predictors $\bm{x}$ and $\bm{w}$ for the response $Y$.
BMA applied to this world results in model averaging across $2^{p}$ potential models, with individual predictors in $\bm{x}$ and $\bm{w}$ either active or not.

The first approach makes use of information only on purchased predictors.
The second approach makes use of information on both purchased and unpurchased predictors.
Information on the unpurchased predictors is available through the conditional distribution of the unpurchased predictors given the purchased (and maybe less expensive) predictors.
It can be considered as an extreme of imputation in the missing value problem, where all cases are missing for some predictors (see also \citet{boone-etal-11}).
The measurement error model also has the similar structure, in that the true value is unobserved (see also \citet{zhang-etal-19} and \citet{doppelhofer-etal-16}).

The predictive loss for both approaches is
\begin{align*}
E \left[ \left\{ Y - h \left( \bm{x}, \bm{w} \right) \right\}^{2} \mid \bm{x} \right],
\end{align*}
where $h ( \cdot )$ is the action as a function of potential predictors.
When the normal linear regression model is used, this loss corresponds to a Bayesian version of the Mallows $C_{p}$ (see \citet{mallows-73}).

The restricted approach removes $\bm{w}$ from the problem, restricting $h$ to be a function of $\bm{x}$ alone, and it averages over a reduced set of models.
This approach leads to the following expression of the loss,
\begin{align*}
E \left[ \left\{ Y - \sum_{\gamma \in \Gamma} h \left( \bm{x}_{\gamma} \right) \pi \left( \gamma \right) \right\}^{2} \mid \bm{x} \right],
\end{align*}
where $\bm{x}_{\gamma}$ is a subset of purchased predictors.

The extended approach marginalizes the loss over $\bm{w}$ by its conditional distribution $g (\bm{w} \mid \bm{x})$, leading to
\begin{align*}
E \left[ \left\{ Y - \sum_{\gamma \in \Gamma, \lambda \in \Lambda} \left( \int h \left( \bm{x}_{\gamma}, \bm{w}_{\lambda} \right) g \left( \bm{w}_{\lambda} \mid \bm{x}_{\gamma} \right) d \bm{w}_{\lambda} \right) \pi \left( \gamma, \lambda \right) \right\}^{2} \mid \bm{x} \right],
\end{align*}
where
$\bm{w}_{\lambda}$ is a subset of unpurchased predictors indexed by $\lambda$ which is defined in a similar manner to $\gamma$
and
$\Lambda$ is a set of all possible $\lambda$s.
In either approach, the optimal action minimizes the sum of predictive loss and cost of predictors.

Both of these approaches can be implemented with standard computational methods.
The restricted approach is standard BMA based on the purchased predictors.
The extended approach is easiest to follow if we assume to know the joint distribution of potential predictors.
In this case, the unpurchased predictors are merely missing data, to be imputed (distributionally) as we fit our model.
When BMA is accomplished by means of Markov chain Monte Carlo (MCMC), standard methods allow us to draw the missing values in each iterate of the algorithm.
If the conditional distribution of $\bm{w} \mid \bm{x}$ is not fully determined, it follows a probability model governed by hyperparameters, it is merely part of the larger Bayesian model, and MCMC or other techniques can be used to perform model averaging over the full set of $2^{p}$ models.

%
%


\subsection{Choice of approach}
\label{subsec:Choice of approach}


One central question is whether the restricted approach or the extended approach is to be preferred.
Our first take on this question is motivated by the subjective Bayesian viewpoint expressed, for example, in \citet{savage-72}.
He constructs Bayesian methods from the principles of rational behavior.
This leads him to the notion of personal probability, and along with it, the ability to specify a prior distribution on unknown parameters (tied to $Y \mid \bm{\phi}$).
The same argument allows one to specify a prior on on models and a prior on the distribution of $\bm{w} \mid \bm{x}$.
This provides a complete description of uncertainty over models, parameters within a model, and missing predictors.
Coupling this with standard results from decision theory which state that the Bayes risk is the minimum possible risk when the parameter follows a given distribution and that the Bayes rule achieves the Bayes risk (Result 1, p. 159 of \citet{berger-85}), we arrive at the usual Bayesian destination.
In other words, the extended approach is preferred from the subjective view.

The implications of this choice run contrary to mainstream Bayesian practice.
Consider a standard BMA problem where one has a set of $k$ predictors, say $\bm{x}$ and a response $Y$.
The usual practice is to apply BMA to the set of all $2^{k}$ models.
While this may appear to agree with the preceding paragraph, we can certainly envision further unobserved predictors $\bm{w}$ that may well be connected to the response at a low cost.
The extended approach averages over these predictors as well, with the analyst's prior beliefs governing the relationship between $\bm{x}$ and $\bm{w}$ and the extended set of model probabilities.

This leads us to ask why BMA is practiced in its current form.
Objective Bayesian methods provide a counterpoint to the subjective Bayesian perspective.
The typical BMA implementation is far from subjective.
Rather than using elicitation procedures to carefully specify a prior distribution across models and, for each model, a prior distribution over the parameters within the model, one resorts to a rule to determine the prior distribution.
The rule may assign a set probability to each model of a given size, and it may routinely specify the distribution on the parameters given the model.
Popular rules include the conjugate priors on model parameters along with the uniform prior model probability (\citet{raftery-etal-97}), the benchmark prior (\citet{fernandez-etal-01}), and the mixture of $g$ priors (\citet{ley-steel-12}).
See \citet{steel-19} for other choices.

Many of these prior distributions are improper, negating the subjective Bayesian argument.
These prior distributions are not constructed in the careful fashion appropriate for smaller scale problems, and they are not accompanied by the claim that all can be modelled, including unseen $\bm{w}$.
A typical attempt at such a specification of the distribution of $\bm{w} \mid \bm{x}$ would lead to an improper distribution for $\bm{w}$.
To see this, replace $\bm{w}$ with $Y$ and note that the marginal distribution on $Y$ is improper for many objective specifications---in particular, for those in which a regression makes use of a uniform improper prior distribution on the intercept or an improper prior distribution on the error variance.
For unseen $\bm{w}$, we may be left without a distribution, and this precludes use of the extended approach.


In addition to the question of whether the extended approach {\em can} be applied under a chosen version of the Bayesian paradigm, there is a question of whether it {\em should} be used.
The major concerns surround our inability to check aspects of the model for future data---our inability to check the form of $Y \mid (\bm{x}, \bm{w})$ when $\bm{w}$ is unavailable and our inability to check the form of $\bm{w} \mid \bm{x}$---and our inability to consistently estimate the distribution of $\bm{w} \mid \bm{x}$ as future data accrue.
This last implies that, even as the future data set size tends to $\infty$, there will always be some uncertainty about the value of observing $\bm{w}$.

The robustness to priors is also an issue when comparing approaches.
The conditional distribution $\bm{w} \mid \bm{x}$ is an additional (subjective) prior.
This specification may or may not be correct, bringing additional sensitivity to the analysis.
If it is based on scientific theory, it helps us to obtain accurate prediction at a lower cost.
However, especially in the area of social science, it is unstable due to, for example, the advance of technology or the change of laws.
Such a misspecification may contaminate the inference, as shown by examples provided in Section 2.1 of \citet{liu-etal-09} and \citet{hahn-19}.
They compares two models with and without biased samples, and discusses that the benefit from the former is smaller than its cost, which in turn suggests the restricted approach over the extended approach from an objective perspective.

\begin{figure}[ht]
\centering
\includegraphics[width=8cm, clip, keepaspectratio]{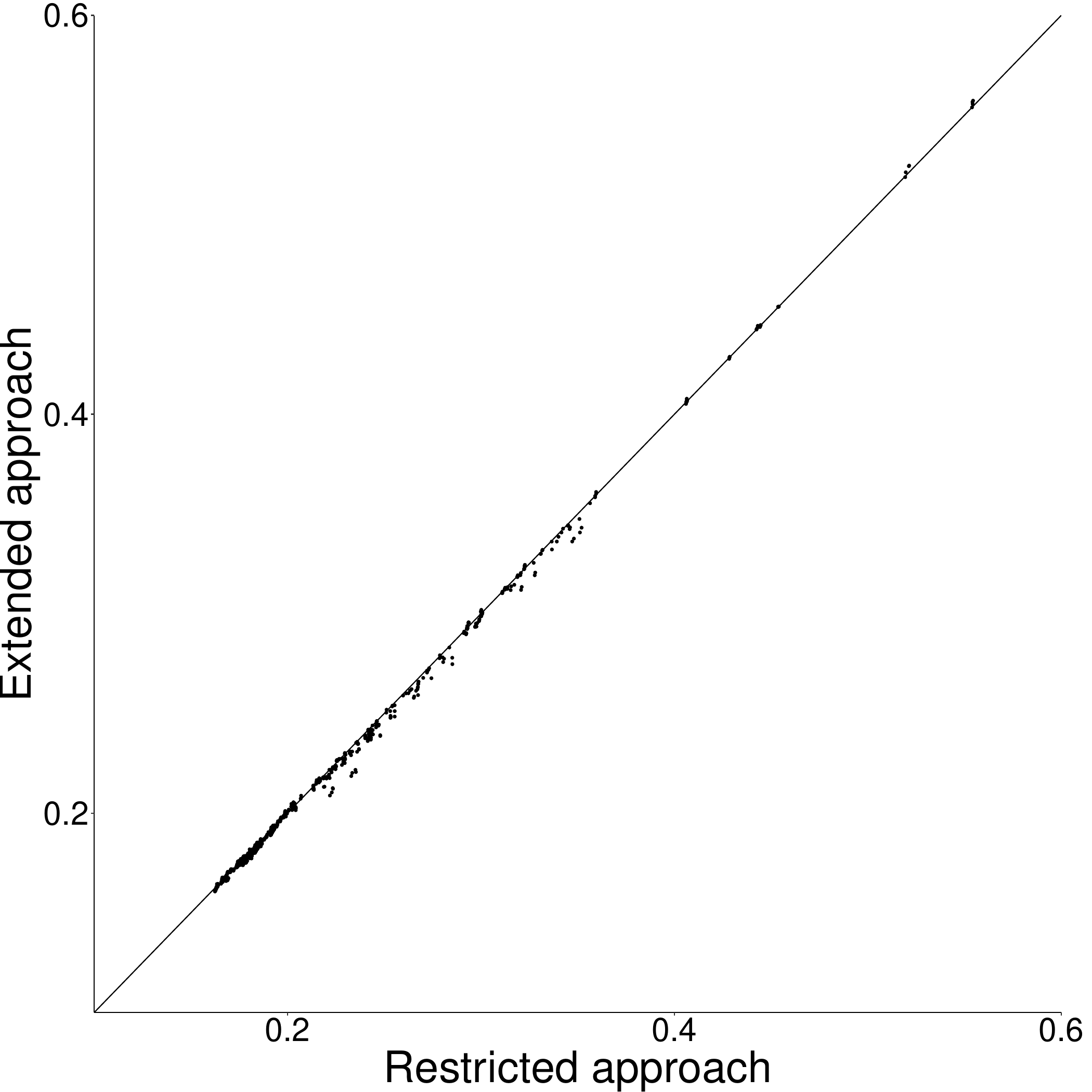}
\caption{Loss plot of two approaches.}
\label{fig:Loss plot of two approaches.}
\end{figure}

At the end of this section, we compare these two approaches by the empirical dataset.
Figure \ref{fig:Loss plot of two approaches.} shows the empirical difference of these two approaches by using the ozone dataset (see Subsection \ref{subsec:Ozone dataset}).
We use the normal linear model with $g$-prior described in Subsection \ref{subsec:Normal linear model with g prior} and draw a pair of losses estimated by taking two approaches (see the next subsection for the loss estimation).
The prior distribution of $\bm{w} \mid \bm{x}$ for the extended approach is constructed by using the normal approximation: use the entire dataset to construct the multivariate normal distribution for $(\bm{x}, \bm{w})$ and derive conditional distributions of unpurchased predictors given purchased ones.
This figure shows these two approaches result in similar losses because they are gathering around the 45 degree line.
Thus, in addition to points listed above as well as the computational aspect, we recommend the restrictive approach, and the paper focuses on it hereafter.


\subsection{Cross-Validated loss}
\label{subsec:Cross-Validated loss}


When the data are observed, we are able to estimate the loss.
Let $(y_{i}, \bm{x}_{i})$ be the response to be predicted and the purchased predictors for case $i$ ($i = 1, \dots, n$), respectively.
Let $D_{\gamma} = \{ y_{i}, \bm{x}_{i, \gamma} \}_{i = 1}^{n}$ denotes the data for each model $\gamma$.

The uncertainties about models and parameters are estimated by the posterior distributions of models and parameters.
%
Then, the loss is estimated as 
\begin{align*}
\left\{ \tilde{y} - \sum_{\gamma \in \Gamma} h \left( \tilde{\bm{x}}_{\gamma} \right) \pi \left( \gamma \mid D_{\gamma} \right) \right\}^{2},
\end{align*}
where $(\tilde{y}, \tilde{\bm{x}}_{\gamma})$ is the new response and predictors for the subset $\gamma$,
$\pi \left( \gamma \mid D_{\gamma} \right)$ is the posterior model probability,
and
$h ( \cdot )$ is an action to be chosen.
Under the squared-error loss, the best action is the posterior conditional expectation $E (\tilde{Y} \mid \tilde{\bm{x}}_{\gamma}, D_{\gamma})$.
%
%

When the new data are not available, the cross-validated loss is an alternative.
The data are split into two parts: the training and validation data.
Then, the conditional expectation and the posterior distributions are estimated based on the training data, and by using the validation data in place of the new data, we have the estimated loss.
When the data are divided into several groups, this process is repeated by treating one of them as the validation and remainings as the training.
The cross-validated loss is the average of these losses.
The remaining of this paper applies the 10-fold cross validation to estimate the predictive loss.
Equation \eqref{eq:spe given model} in Appendix \ref{sec:Posterior, marginal likelihood, and loss} provides the analytical form of loss under the normal linear model with $g$-prior.


\subsection{Prior sensitivity}
\label{subsec:Prior sensitivity}


The next section will illustrate our methodology by using the real datasets.
The results may depend on the prior specification we choose.
This subsection examines the sensitivity from this possibility by using the ozone data to be used in Subsection \ref{subsec:Ozone dataset}.

Taking the restrictive approach, the normal linear models with four prior specifications are considered: (i) $g = k^{2}$, (ii) the hyper-$g$ prior
\begin{align*}
\pi \left( g \right) = \frac{1}{2} \left( 1 + g \right)^{-3/2}, \quad (g > 0),
\end{align*}
(iii) the hyper-$g/n$ prior
\begin{align*}
\pi \left( g \right) = \frac{1}{2n} \left( 1 + \frac{g}{n} \right)^{-3/2}, \quad (g > 0),
\end{align*}
and (iv) the robust prior
\begin{align*}
\pi \left( g \right) = \frac{1}{2} \sqrt{ \frac{1 + n}{1 + k} } \left( 1 + \frac{g}{n} \right)^{-3/2}, \quad (g > (1 + k)^{-1}(1 + n) - 1),
\end{align*}
in addition to the $g$ prior for the last three specifications.

\begin{figure}[!ht]
\begin{minipage}[b]{.48\linewidth}
\centering
\includegraphics[width=6.5cm, clip, keepaspectratio]{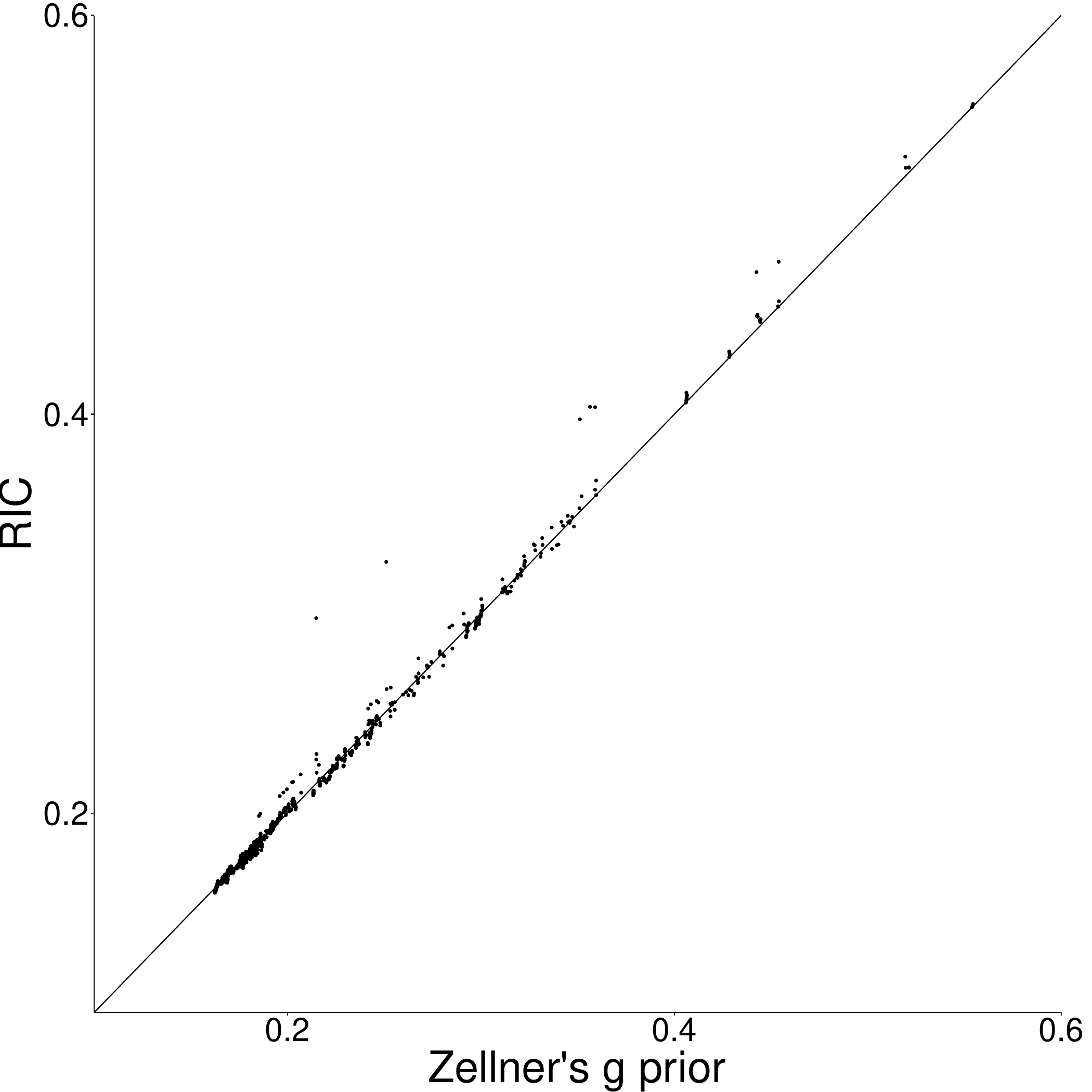}
\subcaption{$g = p^{2}$.}
\end{minipage}
\begin{minipage}[b]{.48\linewidth}
\centering
\includegraphics[width=6.5cm, clip, keepaspectratio]{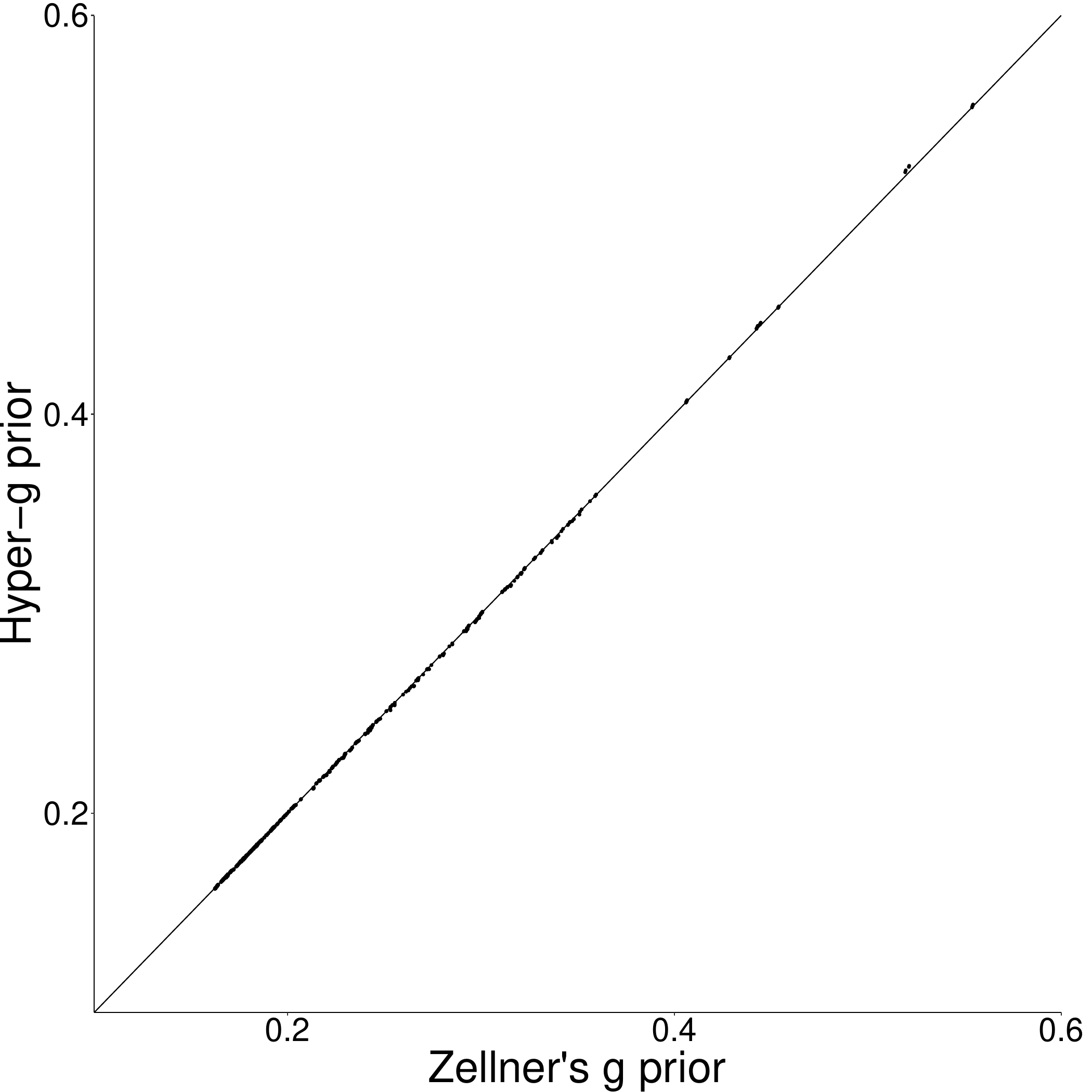}
\subcaption{Hyper-$g$ prior.}
\end{minipage}
\\
\begin{minipage}[b]{.48\linewidth}
\centering
\includegraphics[width=6.5cm, clip, keepaspectratio]{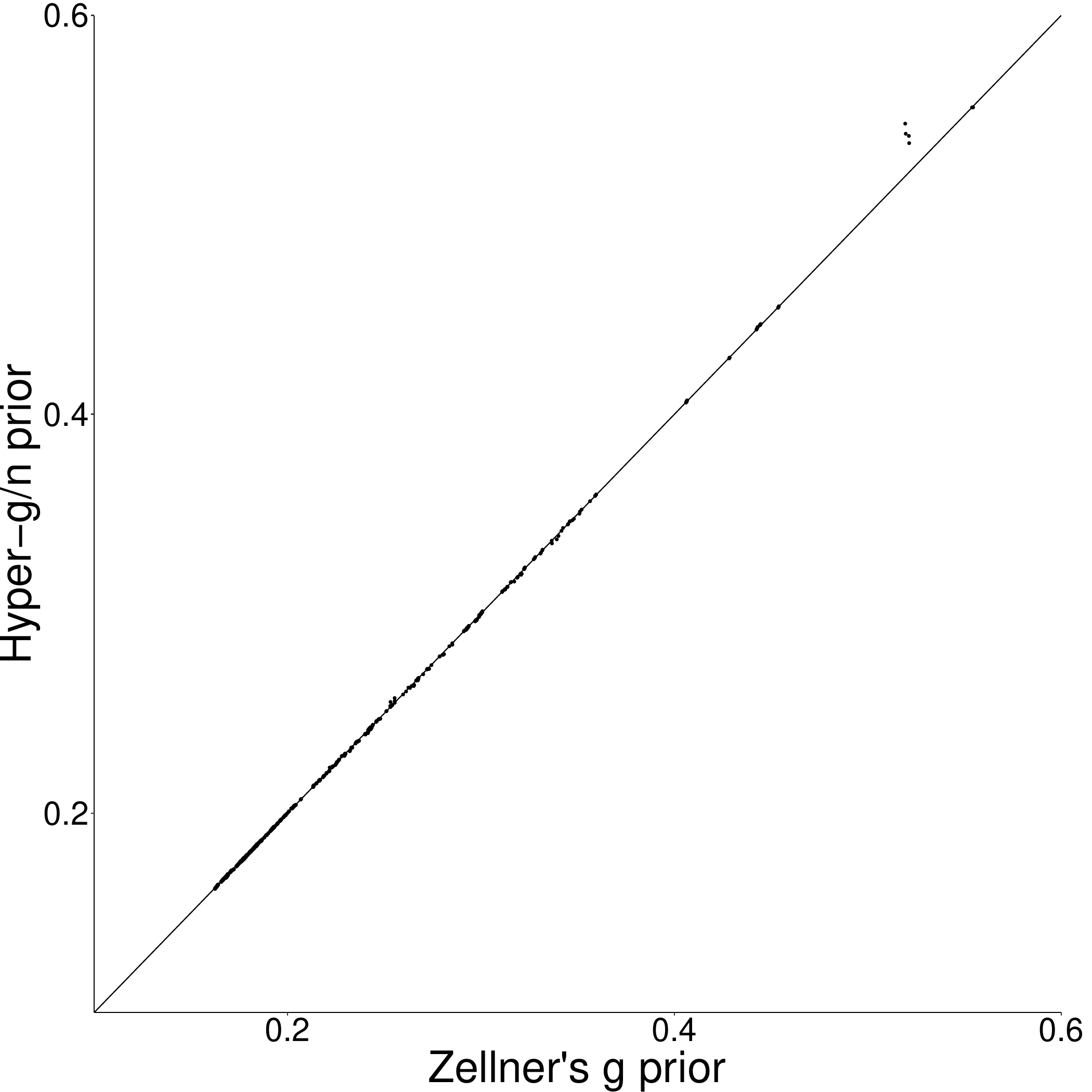}
\subcaption{Hyper-$g/n$ prior.}
\end{minipage}
\begin{minipage}[b]{.48\linewidth}
\centering
\includegraphics[width=6.5cm, clip, keepaspectratio]{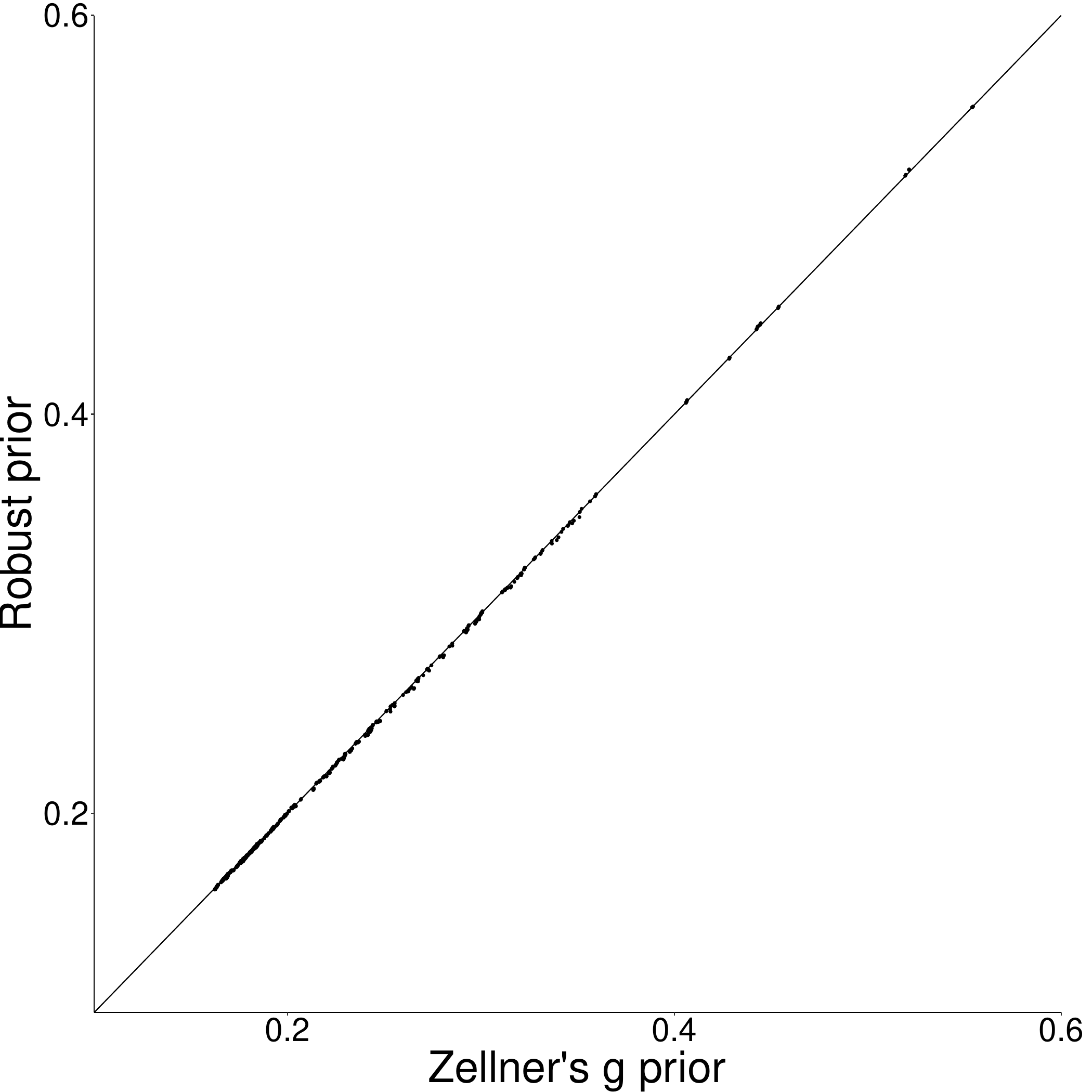}
\subcaption{Robust prior.}
\end{minipage}
\caption{Loss plots of four different prior specifications.}
\label{fig:Loss plots of four different prior specifications.}
\end{figure}

Figure \ref{fig:Loss plots of four different prior specifications.} draws a pair of losses based on different prior specifications, along with the 45 degree line.
For losses under the hyper-$g/n$ prior, the standard Laplace approximation is used as suggested by \citet{liang-etal-08}.
All panels show these five specifications do not have any substantial difference in terms of the squared predictive loss.
Thus, we use the $g$ prior with $g = \max \{ n, k^{2} \}$ as recommended by \citet{fernandez-etal-01} for operational simplicity in the following illustrative examples.


\section{Illustrative examples}
\label{sec:Illustrative examples}


This section illustrates the economic variable selection with three real datasets.
All predictors are standardized and the loss is estimated by using the 10-fold cross validation.


\subsection{Ozone dataset}
\label{subsec:Ozone dataset}


The first dataset is originally analyzed by \citet{breiman-friedman-85} to develop a model between the daily ozone concentration level and meteorological variables in Los Angeles.
We use the data provided by the R package `bfp'.
The number of observations is 330.

The response is the log daily ozone concentration level in 1976 measured at Upland, California.
There are 10 possible predictors: (1) 500-millibar-pressure height, (2) wind speed, (3) relative humidity, (4) temperature at Sandberg, (5) inversion base height, (6) binary variable that is set one if the inversion base height is 5,000, (7) pressure gradient from Los Angeles International Airport to Daggett, (8) inversion base temperature, (9) square root of visibility, and (10) day of year.

\begin{figure}[ht]
\begin{minipage}[b]{.48\linewidth}
\centering
\includegraphics[width=6.8cm, clip, keepaspectratio]{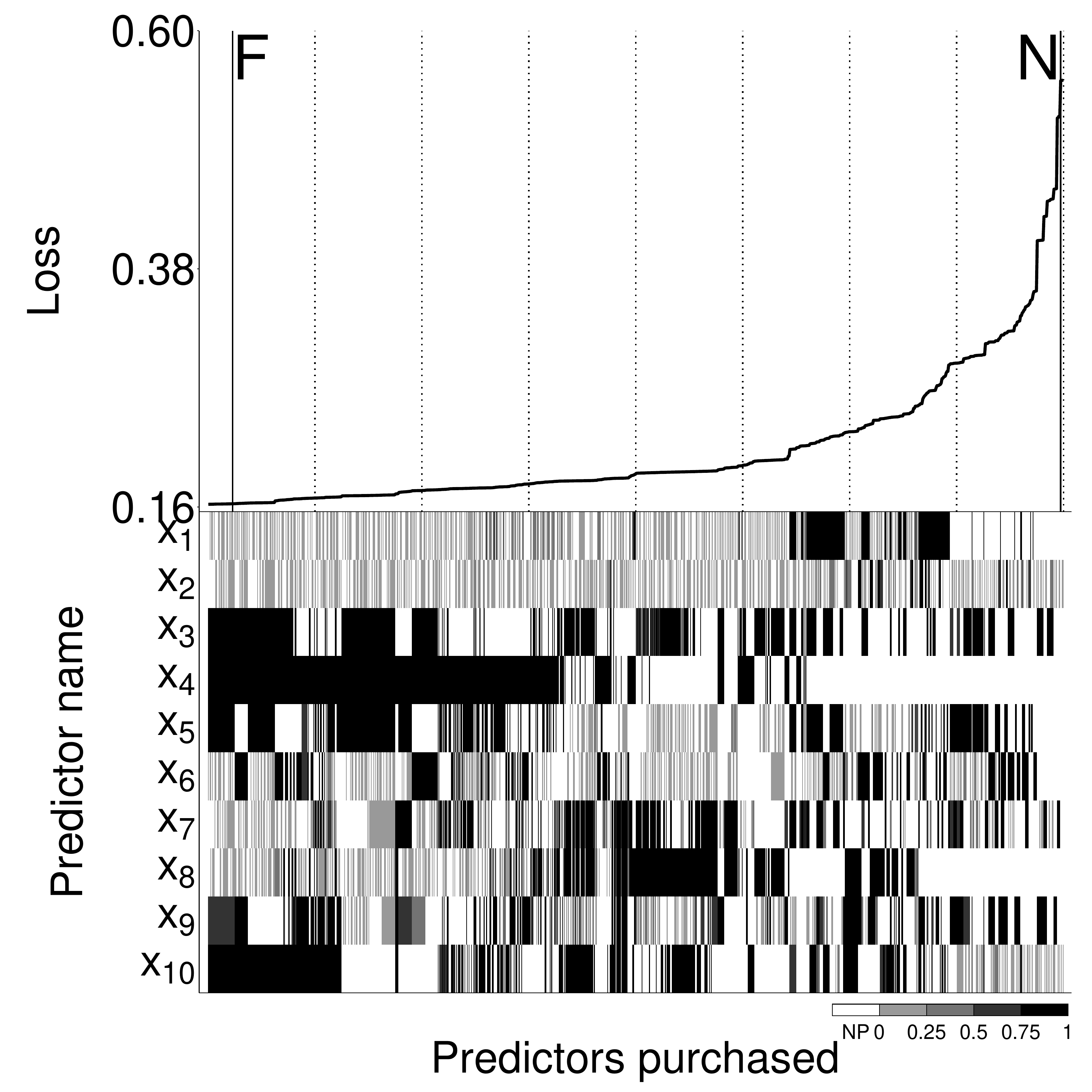}
\subcaption{All combinations.}
\end{minipage}
\begin{minipage}[b]{.48\linewidth}
\centering
\includegraphics[width=6.8cm, clip, keepaspectratio]{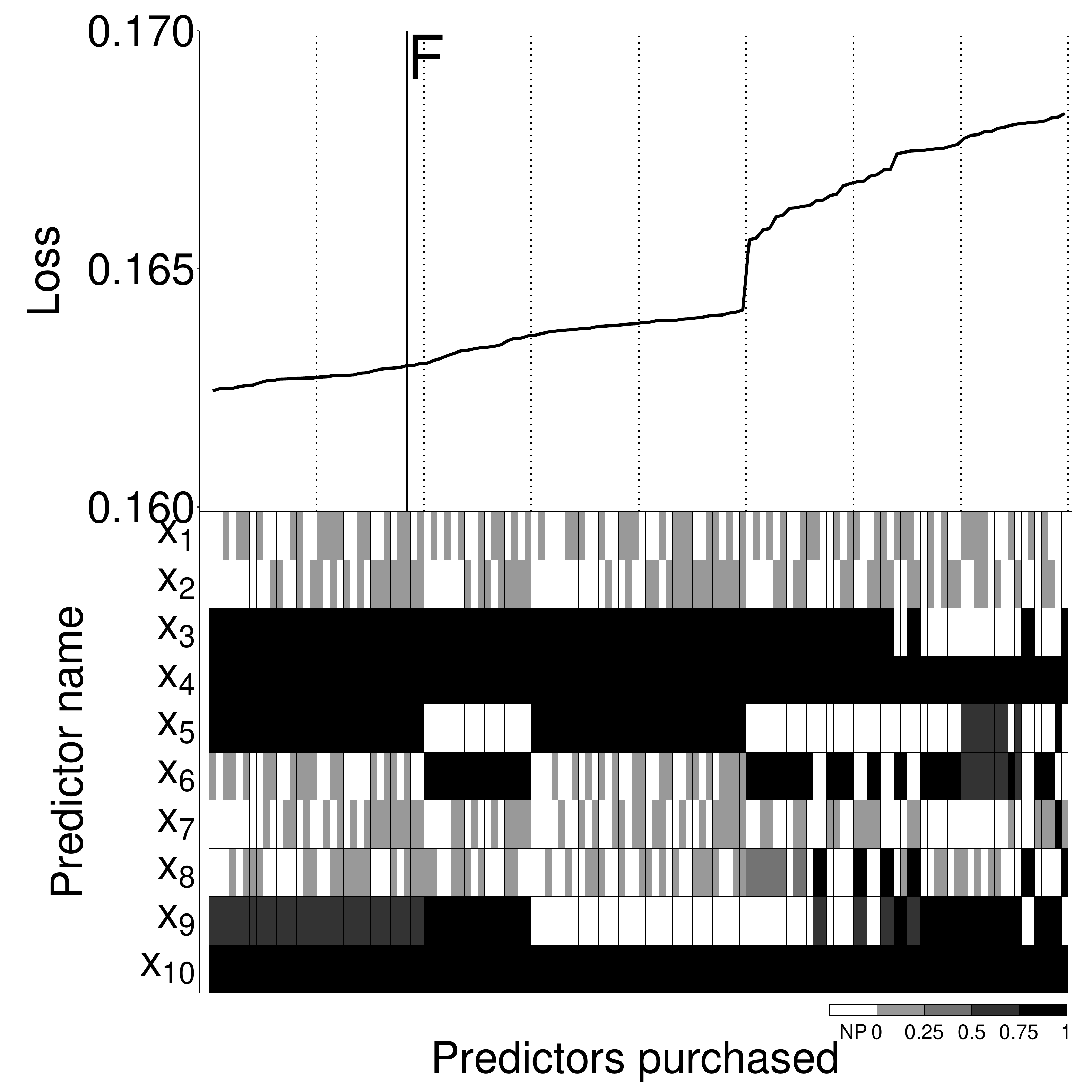}
\subcaption{Top 128.}
\end{minipage}
\caption{Ozone data: selection map with loss plot. F and N denote combinations that purchase all predictors and no predictor, respectively.}
\label{fig:Ozone data}
\end{figure}

Figure \ref{fig:Ozone data} summarizes results.
Both panels consist of two parts: the upper part is the (estimated) squared predictive loss plot in ascending order and each column of the lower part represents a corresponding combination of predictors purchased.
When a cell of a column in the lower part is filled by black, the predictor labeled on the $y$-axis is purchased.
When, on the other hand, it is white, the corresponding predictor is not purchased, which is denoted by NP in the legend provided under the panel.
The marginal posterior probability that the coefficient is nonzero is discretized by the four intervals: $[0, 0.25], (0.25, 0.5], (0.5, 0.75], (0.75, 1]$, and is expressed by the brightness of the cell as shown in the legend.
See \citet{clyde-03} for the marginal posterior nonzero probability.


The least-loss combination is ($x_{3}, x_{4}, x_{5}, x_{6}, x_{9}, x_{10}$).
Among them, $x_{6}$ and $x_{9}$ are less relevant in terms of their marginal posterior nonzero probability (less than 0.6).
Thus, cells corresponding to these predictors are colored to be light gray.
Compared with the selection by \citet{breiman-friedman-85}, we choose $x_{3}$ (and $x_{6}$) instead of $x_{7}$.

This difference is partly due to the transformation of variables.
The response is logged in our example, while it is not in \citet{breiman-friedman-85} (see their Figure 5(a) on page 588).
Among predictors, $x_{7}$ is transformed by a highly nonlinear function in \citet{breiman-friedman-85} (see their Figure 5(d) on page 588), while we do not transform it.
These transformation would be a source of such a difference between their result and ours.

To focus on combinations that yield smaller predictive losses, the left panel is magnified to the right by picking up the top 128 combinations of predictors purchased.
In this panel, ($x_{4}, x_{10}$) are always included in them in terms of its nonzero probability.
Among others, $x_{3}$ is in the combinations with smaller predictive losses.

Two special combinations are considered: the intercept-only combination and the combination that purchases all predictors.
Their respective position is denoted by the vertical solid lines labeled by $N$ and $F$ in Figure \ref{fig:Ozone data}.
The former yields the high predictive loss, although it is not the worst (the fourth from the worst).
On the other hand, the latter performs much better.
This loss is achieved when we use the usual BMA.
Its predictive performance is closer to the best (see the predictive loss plot of the left panel).
However, there are combinations that yield low predictive losses and purchase less predictors.

Next, two cost structures are considered.
The first one is the uniform cost structure, where all predictors are set at the same price.
That is, when $k$ predictors are purchased, the total cost is $c \cdot k$, where $c$ is the price.
This structure is used when a decision maker has no information about the cost of predictors.

When the uniform cost structure is applied, predictors purchased are shown by the left panel of Figure \ref{fig:Ozone data, cost}.
\begin{figure}[ht]
\begin{minipage}[b]{.48\linewidth}
\centering
\includegraphics[width=6.7cm, clip, keepaspectratio]{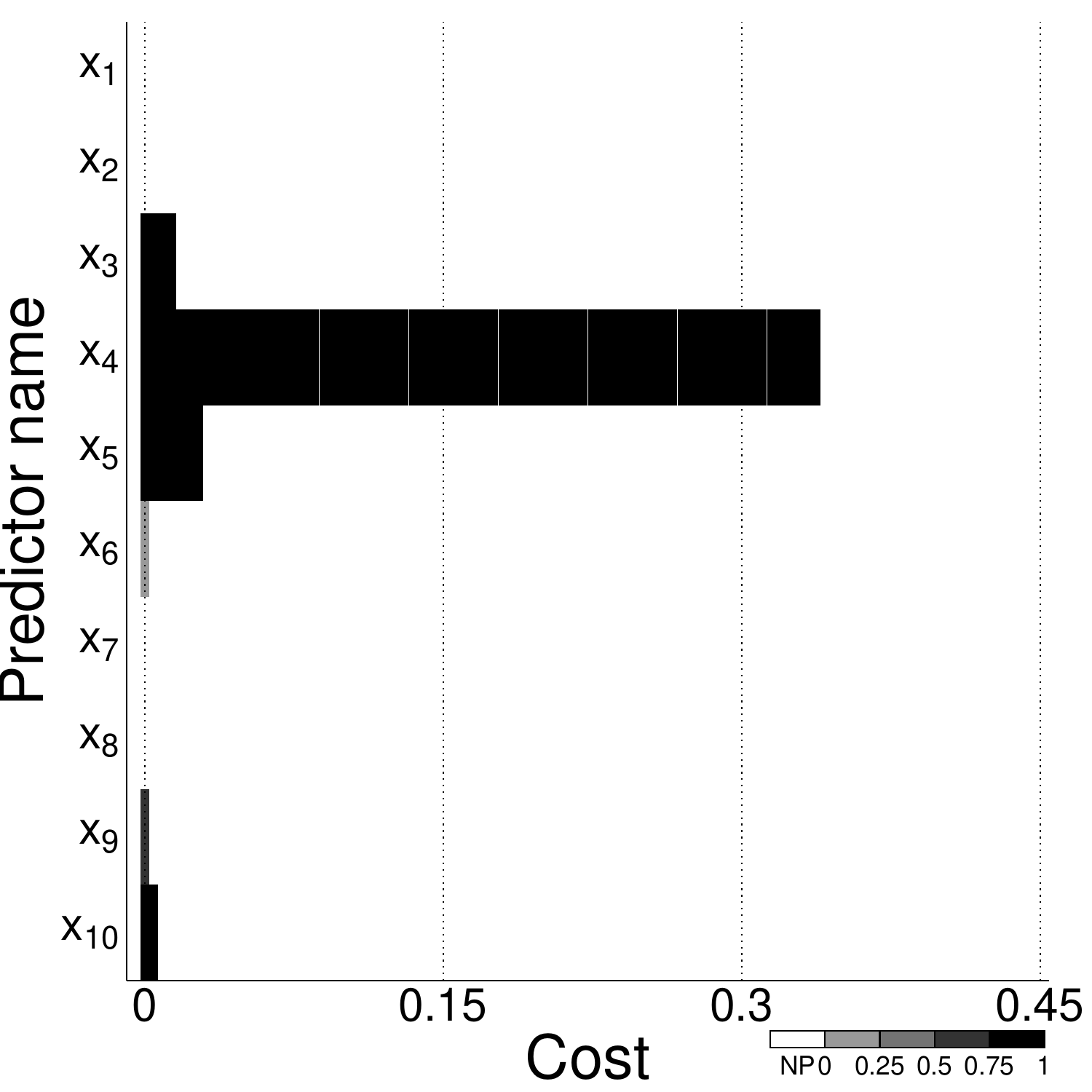}
\subcaption{Uniform cost structure.}
\end{minipage}
\begin{minipage}[b]{.48\linewidth}
\centering
\includegraphics[width=6.7cm, clip, keepaspectratio]{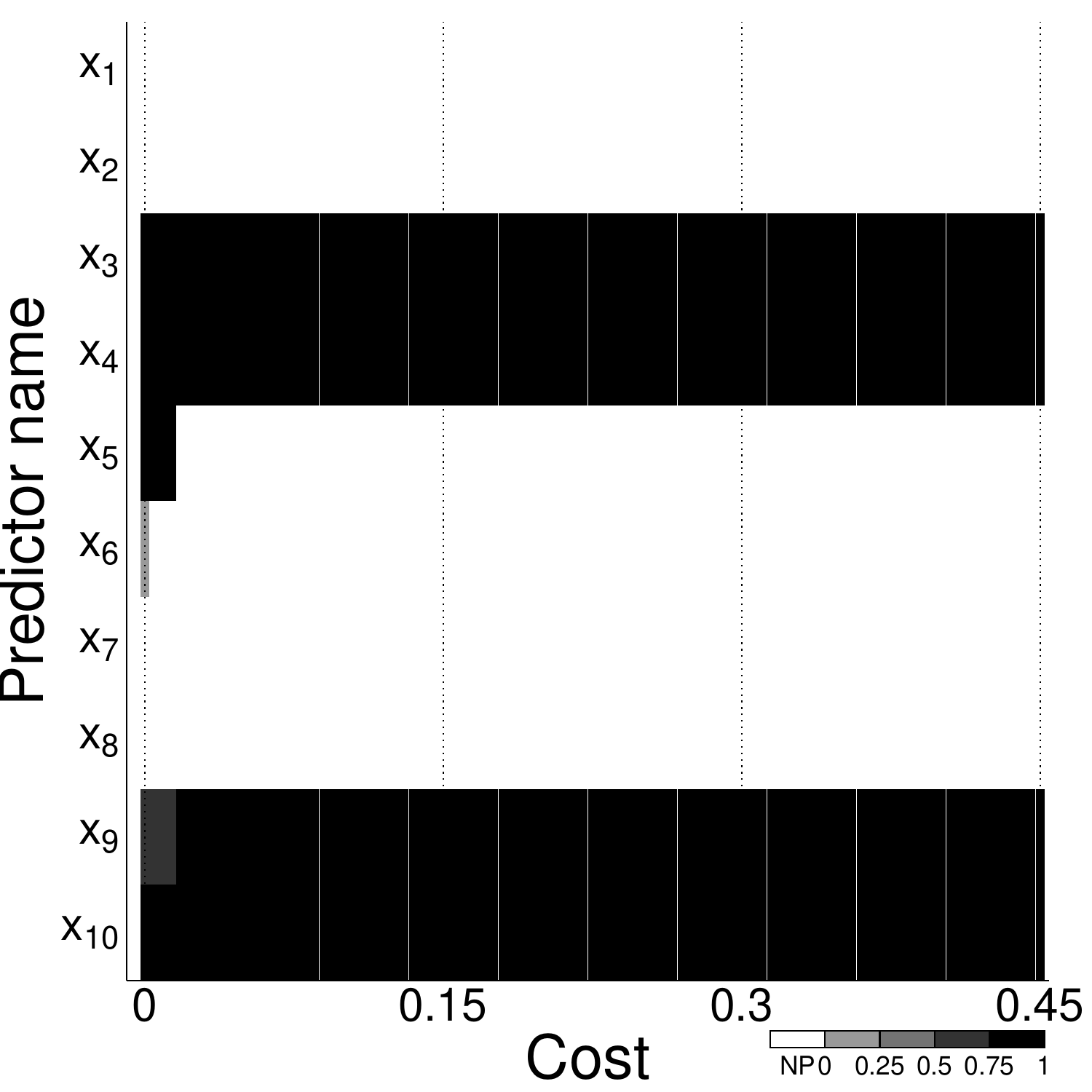}
\subcaption{Cost per predictor.}
\end{minipage}
\caption{Ozone data: least-loss purchases with cost.}
\label{fig:Ozone data, cost}
\end{figure}
Each column is the least-loss combination for a fixed $c$, which is indicated by the $x$-axis.
Similar to the previous plots, the brightness represents the discretized marginal posterior probability that the corresponding coefficient is nonzero.
As $c$ increases, less predictors are purchased.
When $c$ is sufficiently high, the optimal purchase is the one with no predictors.

It is reasonable to consider that a decision is made with some knowledge about the cost of predictors.
A possible decision maker for this dataset is a researcher who is interested in the global warming.
As a part of his or her interest, the researcher would like to predict the ozone level.
He or she probably knows the cost of predictors.
One reasonable cost structure for the researcher is cost per predictor.
Because $(x_{2}, x_{3}, x_{4}, x_{9}, x_{10})$ are often reported on regular weather news, it is natural to assume their costs are zero, while remaining predictors require positive prices.
To simplify the structure, we assume each of them requires the constant price $c$.
The results are shown in the right panel of Figure \ref{fig:Ozone data, cost}.
As $c$ increases, the optimal purchase is the one without $x_{5}$ and $x_{6}$ because of their higher cost.


\subsection{Diabetes dataset}


Next dataset is the diabetes data, which are used in \citet{efron-etal-04} and are provided through Professor Trevor Hastie's webpage.
The data are used to predict the progression of the disease one year ahead of the baseline when predictors related to patients are collected.
In this dataset, 442 observations are included.

The response is the log of diabetes progression measure.
Ten possible predictors are included: (1) age, (2) sex, (3) body mass index, (4) blood pressure, and 6 blood serum measures.

\begin{figure}[ht]
\begin{minipage}[b]{.48\linewidth}
\centering
\includegraphics[width=6.8cm, clip, keepaspectratio]{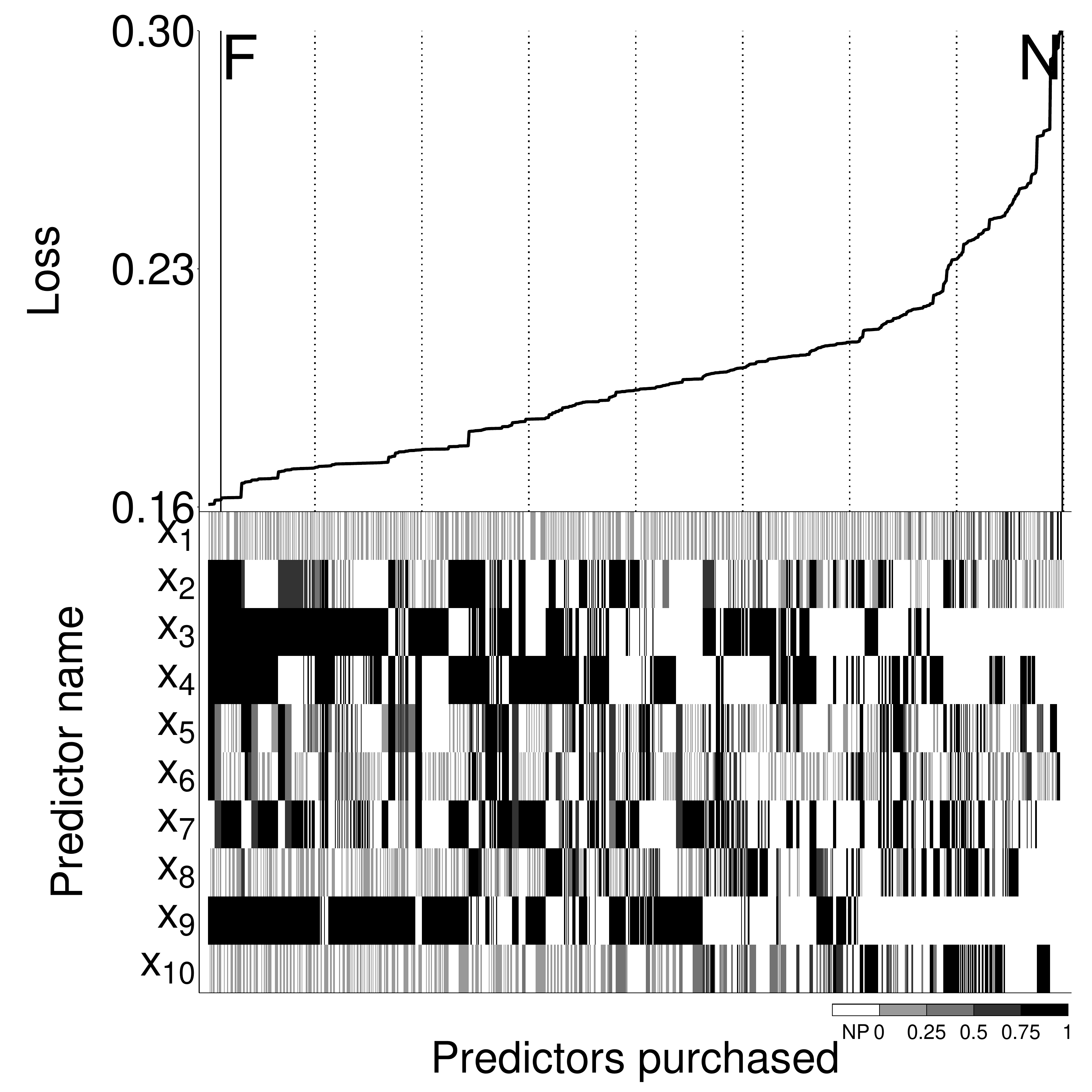}
\subcaption{All combinations.}
\end{minipage}
\begin{minipage}[b]{.48\linewidth}
\centering
\includegraphics[width=6.8cm, clip, keepaspectratio]{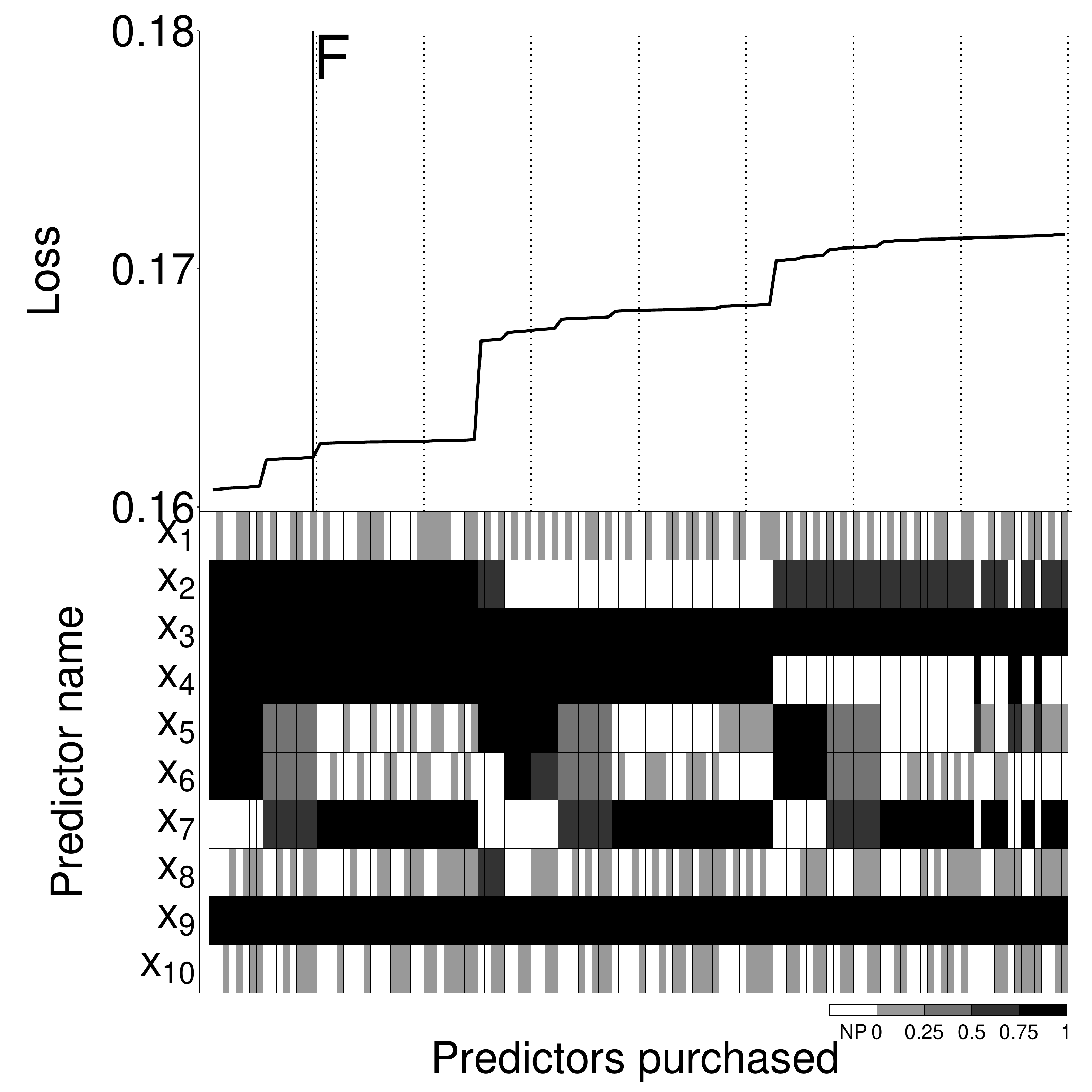}
\subcaption{Top 128.}
\end{minipage}
\caption{Diabetes data: selection map with loss plot. F and N denote combinations that purchase all predictors and no predictor, respectively.}
\label{fig:Diabetes data}
\end{figure}

Results without cost are summarized by Figure \ref{fig:Diabetes data}.
The least-loss combination is $(x_{2}, x_{3}, x_{4}, x_{5}, x_{6}, x_{9})$.
From the top 128 combinations of predictors purchased, ($x_{3}, x_{9}$) are useful to predict the disease progression because it is always included in the combinations.
In addition, $x_{2}$ and $x_{4}$ perform well because they are in low-loss combinations.
Among 6 blood serum measures, $x_{7}$ is useful as well because it is almost always included in the combinations.
\citet{efron-etal-04} applied the least angle regression and they find that variables enter into the active set in the order of $x_{3}, x_{9}, x_{4}, x_{7}$, where they are selected in combinations with higher predictive losses in our results.

The performance of two special combinations are examined.
The intercept-only combination is the second from the worst.
The combination that purchases all predictors does not perform well in this dataset, suggesting purchases that include less predictors would be better to predict the disease progression.

Two specific cost structures are examined.
The first one is the uniform cost structure and its results are on the left panel of Figure \ref{fig:Diabetes data, cost}.
\begin{figure}[ht]
\begin{minipage}[b]{.48\linewidth}
\centering
\includegraphics[width=6.7cm, clip, keepaspectratio]{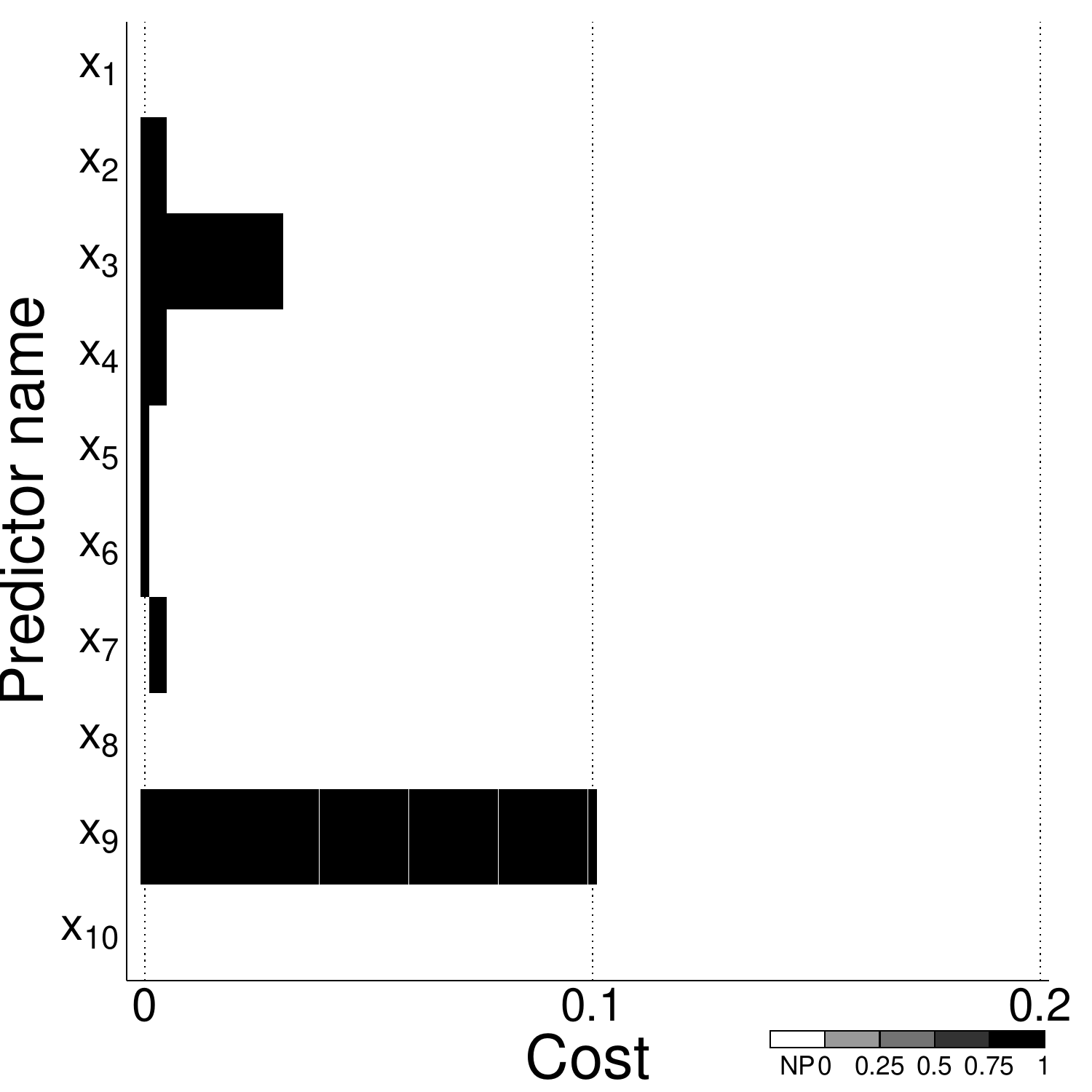}
\subcaption{Uniform cost structure.}
\end{minipage}
\begin{minipage}[b]{.48\linewidth}
\centering
\includegraphics[width=6.7cm, clip, keepaspectratio]{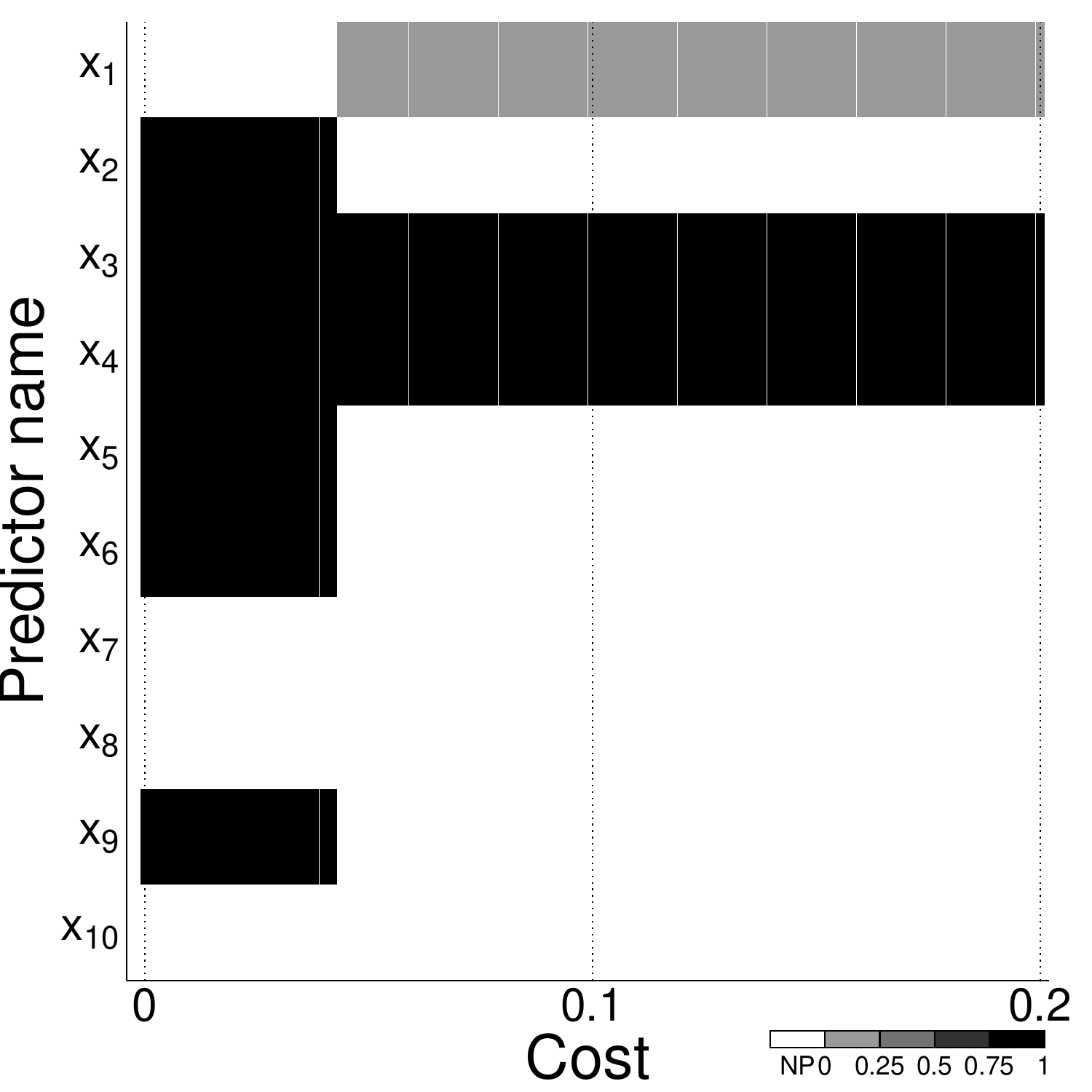}
\subcaption{Cost per model.}
\end{minipage}
\caption{Diabetes data: least-loss purchases with cost.}
\label{fig:Diabetes data, cost}
\end{figure}
As $c$ (the uniform price) increases, the number of predictors in the optimal purchase decreases.
The optimal purchase with sufficiently high price is the one only with the intercept.

A possible decision maker for this dataset is a person who is at the risk of diabetes.
If he or she considers it low, the cost of blood test is expensive.
On the other hand, if he or she considers it high, it becomes cheap.
To simplify this decision problem, the constant cost $c$ is introduced if either of 6 blood serum measures is included in the combination.
Other predictors are assumed to be free of charge.

The results are shown on the right panel of Figure \ref{fig:Diabetes data, cost}.
As the price for the blood test increases, $(x_{5}, x_{6}, x_{9})$ are excluded in the optimal purchase because they become more expensive.
When it is sufficiently high, ($x_{3}, x_{4}$) are selected to predict the progression of the diabetes.
For a person who is at the low risk of diabetes, these predictors are enough for the purpose.


\subsection{Wage dataset}


The last dataset focuses on how wage is determined by attributes of workers, such as the education level and the ability.

The dataset to be used in the analysis is the one taken from the National Longitudinal Survey of Youth and is the panel data from 1979 to 1993.
This is analyzed by \citet{koop-tobias-04} and is provided from the Journal of Applied Econometrics data archive.
The response is the log of hourly wage for white males.
\citet{koop-tobias-04} excluded observations who are at the age of less than 16 years or who report small wages, short working hours, or inappropriate education years.
There are 7 possible predictors: (1) education in years, (2) potential experience in years (age $-$ years of education $-$ 5), (3) the ability measure ranging from about $-4$ to about $2$, constructed on 10 component tests of the Armed Services Vocational Aptitude Battery, (4) mother's education in years, (5) father's education in years, (6) binary variable for broken home until the age of 14, and (7) number of siblings.
The response and the first two variables are time variant, while the remaining five are time invariant.
More details of this dataset are given in Section 4 of \citet{koop-tobias-04}.

\begin{figure}[ht]
\centering
\includegraphics[width=8cm, clip, keepaspectratio]{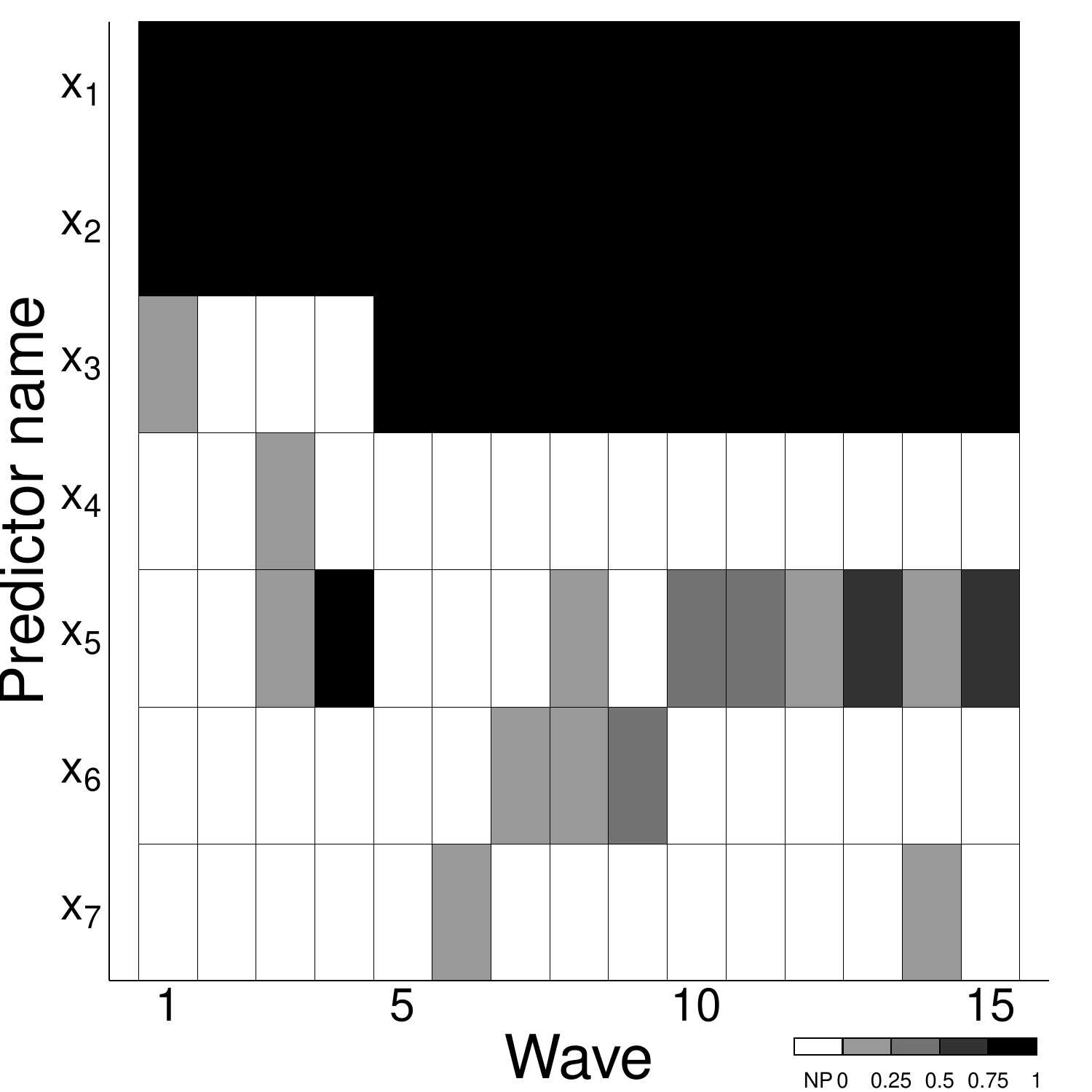}
\caption{Wage data: least-loss combinations by waves.}
\label{fig:Least-loss combinations (wage).}
\end{figure}

The least-loss combination of predictors purchased for each wave is aligned in Figure \ref{fig:Least-loss combinations (wage).}.
The ability measure ($x_{3}$) comes into the set of predictors purchased after the fifth wave in terms of its marginal posterior nonzero probability more than 0.75.
A possible reason is as follows.
For the first four years, companies mainly set wages by the education level ($x_{1}$) and the experience ($x_{2}$) because the ability is unobservable at this moment.
It will be turned out as working together.
After about four years, companies start to use its information to set wages more accurately.

A decision problem in this dataset is when to purchase the ability measure as a manager of a company.
When it is free of charge, purchasing at the beginning of $t$-th wave yields the prediction loss as $\sum_{s = 1}^{t-1} l_{s} + \sum_{s = t}^{15} l_{s}^{\ast}$, where $l_{s}^{\ast}$ and $l_{s}$ are the least losses with and without purchasing the ability measure, respectively.

\begin{figure}[ht]
\begin{minipage}[b]{.48\linewidth}
\centering
\includegraphics[width=6.5cm, clip, keepaspectratio]{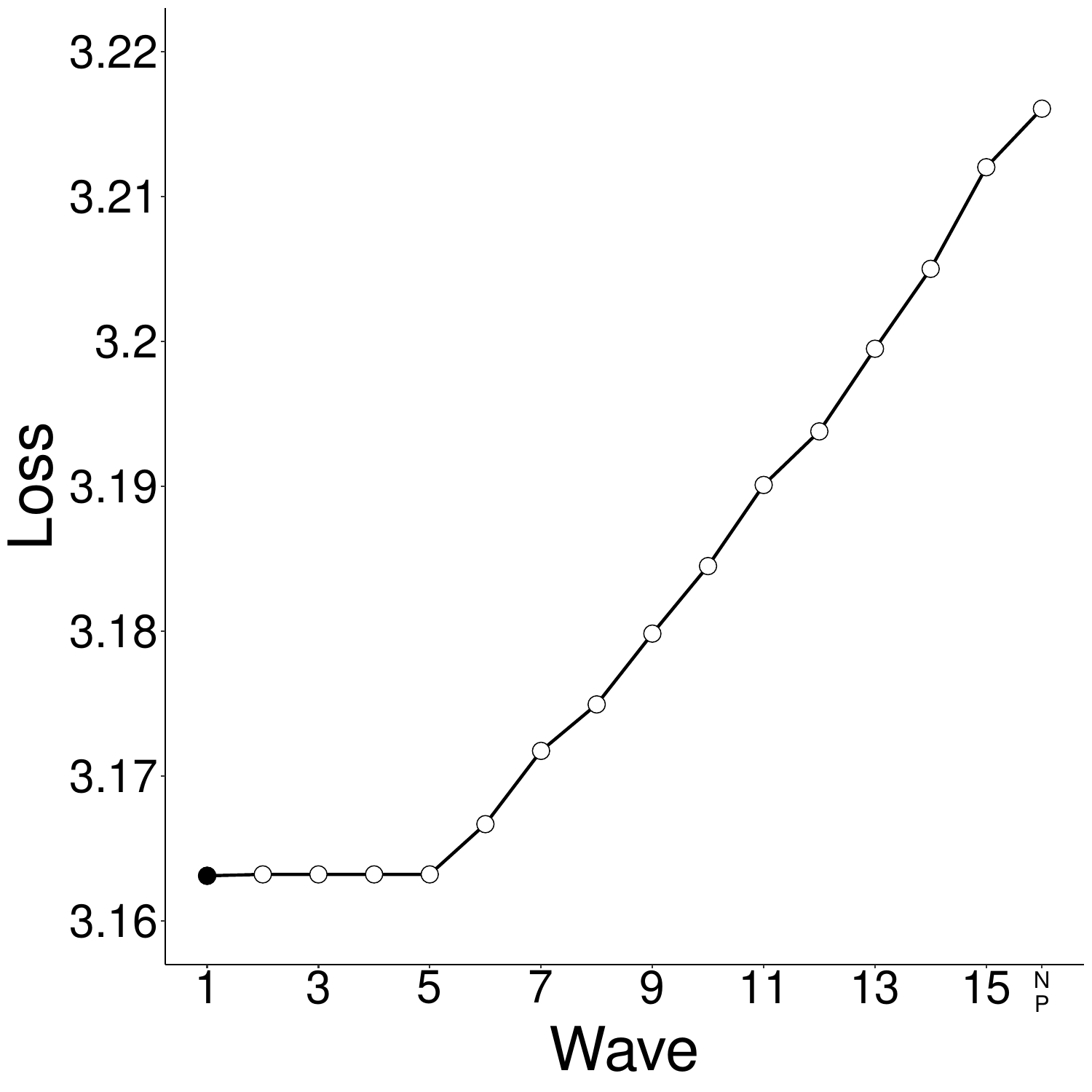}
\subcaption{No cost.}
\end{minipage}
\begin{minipage}[b]{.48\linewidth}
\centering
\includegraphics[width=6.5cm, clip, keepaspectratio]{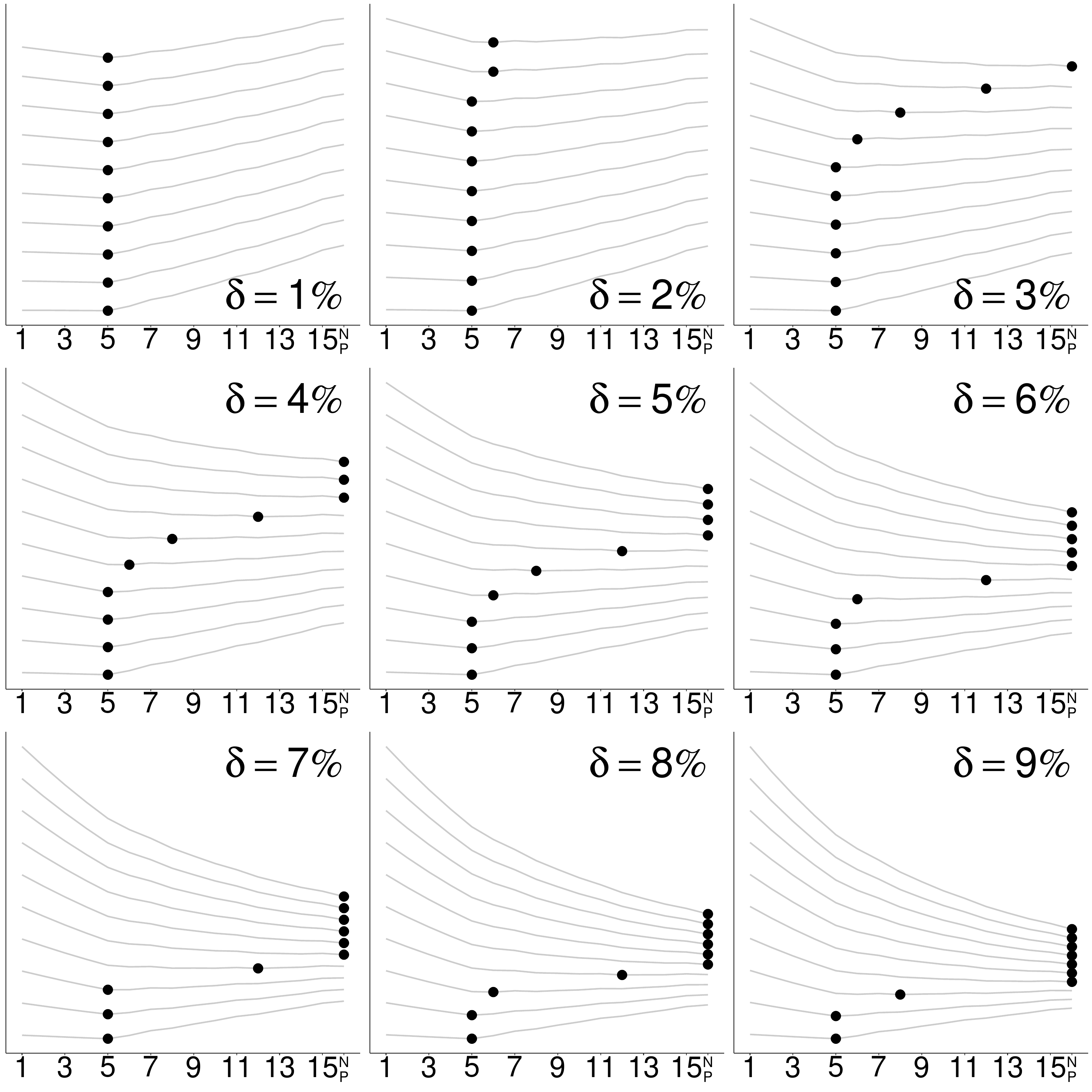}
\subcaption{Cost introduced.}
\end{minipage}
\caption{Wage data: when to purchase. NP denotes the combination that purchases no predictor for all waves.}
\label{fig:When to purchase (wage).}
\end{figure}

The left panel of Figure \ref{fig:When to purchase (wage).} plots this loss changing when to purchase.
%
The minimum loss is reached at the beginning of the first wave (represented by the black dot), although losses at the beginning of the second through fifth waves are comparable.

The decision changes when cost is introduced.
There are two kinds of cost in this problem: the discount factor ($\delta$) and the price of the ability measure ($c$).
Because the present and future utilities/costs are not equivalent, the discount factor is introduced to evaluate the future in terms of the present value.
If the ability measure was purchased at the beginning of the $t$-th wave, the prediction loss with cost adjusted by the discount factor would be
\begin{align*}
\sum_{s = 1}^{t-1} \frac{l_{s}}{\left( 1 + \delta \right)^{s-1}} + \sum_{s = t}^{15} \frac{l_{s}^{\ast}}{\left( 1 + \delta \right)^{s-1}} + \frac{c}{\left( 1 + \delta \right)^{t-1}}.
\end{align*}
The loss is discounted because it is interpreted as the utility in the statistical decision problem or because it is measured by dollars so that it is additive to the cost.

The decision with cost is shown in the right panel of Figure \ref{fig:When to purchase (wage).}.
Each panel is for a fixed discount factor and ten different prices of the ability measure, that are ranging from 0.01 to 0.2.
A gray line is the plot of the loss described above for a fixed price and discount factor, and a black dot indicates the minimum.

As the price gets higher, the loss increases.
When we see the top left panel as an example, the gray line moves up along the $y$-axis as the price increases.
Thus, in this panel, the optimal decision stays at the beginning of the fifth wave.
With the positive discount factor, it moves to a later wave or no purchase as the price increases.
See the top right panel, for example.

As the discount factor becomes larger, the loss plot becomes downward because the decision maker values the present more than the future.
Then, the optimal decision tends to purchase the measure at a later wave even if it improves the loss.
When the discount factor and price are sufficiently large, the optimal decision is no purchase.
It is reasonable that the manager of a company purchases the ability measure later or decides no purchase of it when it is expensive and/or the discount factor is large.


\section{Discussion}
\label{sec:Discussion}


The variable selection problem depends on the person who chooses predictors.
This aspect is formulated as a decision problem, and the optimal decision is the BMA with purchased predictors.
In a broader view, it is considered to be the restricted approach.
The extended approach is another methodology to select predictors when the subjective prior information about the distribution of unpurchased predictors conditional on purchased ones is available.
%
As discussed in Subsection \ref{subsec:Choice of approach}, the restricted approach is our recommendation.

Empirical results that employ the restricted approach show that the predictive loss is improved with a subset of predictors, compared with the one with all predictors.
Cost structures specific to the dataset is also considered.
We find that the optimal decision (the optimal set of predictors to be purchased in this case) changes according to the structure or the level of the price for predictors.


Finally, a computational issue is noted.
The method is computationally feasible when the number of predictors is moderate.
However, for example, the growth model usually includes more than fifty predictors (sixty seven in \citet{sala-i-margin-etal-04}).
In this case, a simple method such as the stepwise selection would be useful to remove less relevant predictors before proceeding to apply our method (see, e.g., \citet{james-etal-13} for the stepwise selection).
However, a decision-theoretic variable selection in high dimensions is an interesting research question and will be left as a future work.




\appendix
\section{Posterior, marginal likelihood, and loss}
\label{sec:Posterior, marginal likelihood, and loss}


This section derives the posterior, the marginal likelihood, and the loss under the normal linear regression model specified in Subsection \ref{subsec:Normal linear model with g prior}.
The subscript $\gamma$ is suppressed in this section to simplify notation, except for the number of predictors.
We use $k$ instead of $p_{\gamma}$.

Suppose we have the training data $D = \{ y_{i}, \bm{x}_{i} \}_{i = 1}^{n}$.
The matrix representation gives $\bm{y} = (y_{1}, \dots, y_{n})^{\prime}$ and $\bm{X} = (\bm{x}_{1}, \dots, \bm{x}_{n})^{\prime}$.
Consider the following normal linear regression and prior distribution:
\begin{align*}
Y_{i} &= \beta_{0} + \bm{x}_{i}^{\prime} \bm{\beta}_{1} + \epsilon_{i},
\quad
\epsilon_{i} \overset{i.i.d.}{\sim} N (0, \sigma^{2}),
\quad
i = 1, \dots, n, \notag \\
\pi \left( \beta_{0} \right) &\propto 1,
\quad
\bm{\beta}_{1} \sim N_{k} \left\{ \bm{0}, g \sigma^{2} \left( \bm{X}^{\prime} \bm{X} \right)^{-1} \right\},
\quad
\pi \left( \sigma^{2} \right) \propto \sigma^{-2}, \notag
\end{align*}
Then, we have an analytical form for the (marginal) posterior distribution of the regression coefficients, $\bm{\beta} = (\beta_{0}, \bm{\beta}_{1}^{\prime})^{\prime}$.
The posterior is the Arellano-Valle and Bolfarine's generalized $t$ distribution, which is given by
\begin{align}
\bm{\beta} \mid D
&\sim
t \left( \bm{b}, \bm{B}; S, n-1 \right),
\label{eq:marginal posterior distribution}
\intertext{where
$\bar{y} = \frac{1}{n} \sum_{i = 1}^{n} y_{i}$,
$d_{y}^{2} = \bm{y}^{\prime} \bm{y} - n \bar{y}^{2}$,
}
\bm{b} &= \begin{pmatrix} \bar{y} \\ \frac{g}{1 + g} \left( \bm{X}^{\prime} \bm{X} \right)^{-1} \bm{X}^{\prime} \bm{y} \end{pmatrix},
\quad
\bm{B} = \begin{pmatrix} \frac{1}{n} & \bm{0}^{\prime} \\ \bm{0} & \frac{g}{1 + g} \left( \bm{X}^{\prime} \bm{X} \right)^{-1} \end{pmatrix}, \notag \\
R^{2} &= \frac{ \bm{y}^{\prime} \bm{X} \left( \bm{X}^{\prime} \bm{X} \right)^{-1} \bm{X}^{\prime} \bm{y} }{ d_{y}^{2} },
\quad
S = \frac{d_{y}^{2}}{1 + g} \left\{ 1 + g \left( 1 - R^{2} \right) \right\}. \notag
\end{align}
The probability density function is given in \citet{arellano-valle-bolfarine-95}.
The posterior expectation and variance matrix of $\bm{\beta}$ are $\bm{b}$ and $\frac{S}{n-3} \bm{B}$, respectively.

The marginal likelihood is derived as
\begin{align}
m \left( \bm{y} \mid \bm{X} \right)
=
\frac{ \Gamma \left( (n-1)/2 \right) }{ \sqrt{\pi}^{n-1} \sqrt{n} }
\left( 1 + g \right)^{(n - k - 1)/2}
d_{y}^{-(n-1)}
\left\{ 1 + g \left( 1 - R^{2} \right) \right\}^{-(n-1)/2},
\label{eq:marginal likelihood}
\end{align}
where $\Gamma (x)$ is the gamma function (see also \citet{steel-19} for this expression).

Finally, the squared predictive loss given the model is estimated by
\begin{align}
%
\frac{1}{m}
\sum_{i = 1}^{m} \left\{ \tilde{y}_{i} - \bar{y} - \frac{g}{1 + g} \tilde{\bm{x}}_{i}^{\prime} \left( \bm{X}^{\prime} \bm{X} \right)^{-1} \bm{X}^{\prime} \bm{y} \right\}^{2},
\label{eq:spe given model}
\end{align}
%
where $(\tilde{y}_{1}, \dots, \tilde{y}_{m})^{\prime}$ and $(\tilde{\bm{x}}_{1}, \dots, \tilde{\bm{x}}_{m})^{\prime}$ are the response and predictors in the validation set with $m$ observations.


\bibliographystyle{chicago}
\bibliography{bma_bib}


\end{document}